# The Discovery of Tunable Universality Class in Superconducting $\beta-\mathrm{W}$ Thin Films

Ce Huang[1,2†], Enze Zhang[1,2†], Yong Zhang[3†], Jinglei Zhang[4†], Faxian Xiu[1,2,5,6*], Haiwen Liu[7*], Xiaoyi Xie[1,2], Linfeng Ai[1,2], Yunkun Yang[1,2], Minhao Zhao[1,2], Junjie Qi[8], Lun Li[3], Shanshan Liu[1,2], Zihan Li[1,2], Runze Zhan[9], Ya-Qing Bie[9], Xufeng Kou[3], Shaozhi Deng[9], X. C. Xie[8,10,11]

[1] State Key Laboratory of Surface Physics and Department of Physics, Fudan University, Shanghai 200433, China

[2] Institute for Nanoelectronic Devices and Quantum Computing, Fudan University, Shanghai 200433, China

[3] School of Information Science and Technology, ShanghaiTech University, Shanghai 201210, China

[4] Anhui Province Key Laboratory of Condensed Matter Physics at Extreme Conditions, High Magnetic Field Laboratory of the Chinese, Academy of Sciences, Hefei, China

[5] Collaborative Innovation Center of Advanced Microstructures, Nanjing 210093, China

[6] Shanghai Research Center for Quantum Sciences, Shanghai 201315, China

[7] Center for Advanced Quantum Studies, Department of Physics, Beijing Normal University, 100875 Beijing, China

[8] Beijing Academy of Quantum Information Sciences, West Bld.3, No.10 Xibeiwang East Rd., Haidian District, Beijing 100193, China

[9] State Key Lab of Optoelectronic Materials and Technologies, Guangdong Province Key Laboratory of Display Material and Technology, School of Electronics and Information Technology, Sun Yat-sen University, Guangzhou, 510275, People's Republic of China.

[10] International Center for Quantum Materials, School of Physics, Peking University, 100871 Beijing, China.

[11] CAS Center for Excellence in Topological Quantum Computation, University of Chinese Academy of Sciences, Beijing 100190, China

[†]These authors contributed equally to this work.

[*]Correspondence and requests for materials should be addressed to F. X. (E-mail: Faxian@fudan.edu.cn) and H. L. (E-mail: Haiwen.liu@bnu.edu.cn)

**Abstract: The interplay between quenched disorder and critical behavior in quantum phase transitions is conceptually fascinating and of fundamental importance for understanding phase transitions. However, it is still unclear whether or not the quenched disorder influences the universality class of quantum phase transitions. More crucially, the absence of superconducting-metal transitions under in-plane magnetic fields in 2D superconductors imposes constraints on the universality of quantum criticality. Here, we discover the tunable universality class of superconductor-metal transition by changing the disorder strength in $\beta-\mathrm{W}$ films with varying thickness. The finite-size scaling uncovers the switch of universality class: quantum Griffiths singularity to multiple quantum criticality at a critical thickness of $t_{c\perp1}\sim8$ nm and then from multiple quantum criticality to single criticality at $t_{c\perp2}\sim16$ nm. Moreover, the superconducting-metal transition is observed for the first time under in-plane magnetic fields and the universality class is changed at $t_{c\parallel}\sim8$ nm. The discovery**

**of tunable universality class under both out-of-plane and in-plane magnetic fields provides broad information for the disorder effect on superconducting-metal transitions and quantum criticality.**

Phase transitions belong to the most fascinating phenomena in nature. Quantum fluctuation effect[1] can give rise to quantum phase transitions (QPTs)[2] such as the superconductor-metal transition (SMT) in low-dimensional superconductors, the transition between quantized Hall plateaus and non-Fermi-liquid phases in a Kondo lattice[1-9]. The ordered and disordered phases in QPTs are separated by the quantum critical point (QCP) at zero temperature and the physics behind is uncovered by quantum criticality. The simultaneous interplay of quantum fluctuation and dissipation greatly enriches the physical characteristics of QPT around the QCP[1,4,6,9-12], giving rise to unprecedented quantum phenomena, such as the quantum Griffiths singularity (QGS) in two dimensional (2D) and quasi-1D superconductors[13-19], preformed Cooper pairs in disordered superconductors[20,21] quantum metal state[22,23], quantum spin liquid[24], and the change of QCP position by pressure[25].

The quenched disorder can induce pronounced quantum fluctuation and is generally believed to be the key to QPTs. The role of disorder and dissipation on SMT has been considered within large theoretical frameworks[4,26-32], which provide various predictions on the QPT behaviors. Despite the number of experiments followed up with a large variety of correlation length critical exponents ($\nu$) that have been performed in various systems[4], unfortunately, the inconsistency among the critical exponents found in different physical systems is a source of great controversy and that increases the difficulty of further investigation. Besides, it is unknown whether and how disorder can tune the dynamic critical exponent $z$ which has been predicted to depend on models with random interactions[33] and even diverge in a random transverse field Ising model[7]. The different $z$ values determine the scaling law and the corresponding universality class. Thus, a better understanding of the relationship between disorder and universality class is highly desirable which is both conceptually interesting and of fundamental importance. The tunable universality class (TUC) emerges as direct evidence to resolve the problem of the QPTs and provide the universal conclusion between various critical behaviors and disorder strength. Apparently, to date, there has not been any material systems that exhibit TUC. More critically, the absence of the SMT behavior under in-plane magnetic fields in 2D superconductors also imposes constraints on the universality of quantum criticality.

Here, we systematically study the critical behaviors in thin superconducting $\beta-$W films. By changing the film thickness and simultaneously varying the disorder strength, we discover the TUC with entirely different critical behaviors. The notable TUC features are revealed under both out-of-plane and the in-plane magnetic fields. Specifically, the universality class of SMT switches twice under out-of-plane magnetic fields when increasing the thickness: from QGS to multiple quantum criticality and then from multiple quantum criticality to single criticality with a critical thickness of 8 nm and 16 nm, respectively. Importantly, the scaling procedure gives a critical exponent $z\nu\sim30$ in a 2 nm-thick film at 0.05 Kelvin. And, to our knowledge, it is the highest value ever reported. Meanwhile, the SMT is observed for the first time under in-plane magnetic fields, which also demonstrates a similar TUC behavior. The discovery of TUC provides an overall and self-consistent framework for the disorder effect on SMT and quantum criticality, which deepens the understanding of QPT driven by disorder and fluctuation.

$\beta-\mathrm{W}$ crystalizes in a close-packed A15 structure with a $Pm\bar{3}n$ space group[34] as depicted in Fig. 1(a) and exhibits $s$-wave superconductivity behavior with $T_c \sim 3$ K[35, 36]. Here, uniform $\beta-\mathrm{W}$ films with various thicknesses were grown by magnetron sputtering deposition with 10 nm thick $SiO_2$ films deposited on top as a protection layer for *ex-situ* transport measurements (see Supplementary Section 1 and Fig. S1-2 for details). Standard four-probe measurements of $\beta-\mathrm{W}$ films were carried out in Fig. 1b inset. The superconductivity transition, defined at 90% of the normal state resistance $0.9R_n$, occurs at $T_c = 1.64$ K in Fig. 1(b) for a 2 nm-thick sample which has the largest disorder strength (Table S1 and Fig. S3). The 2D superconductivity is confirmed by BKT phenomenology[37,38] in zero field temperature-dependent resistance (Fig. S4(c)), the angular dependence of the critical fields[39] (Fig. S4(a)-(b)) and the criteria of $t < \xi_{GL}(0)$, where $t$ and $\xi_{GL}(0)$ are thickness and Ginzburg-Landau (GL) coherence length[40] (Fig. S5-6 and Supplementary Section 4).

The upper critical field ($\mu_0 H_{c2}$) of the 2 nm-thick $\beta-\mathrm{W}$ films is plotted in Fig. 1(c) (see the original temperature-dependent magnetoresistance data in Fig. S5(a)-(b)). Under out-of-plane magnetic fields, $\mu_0 H_{c2}^{\perp}$ deviates from the Werthamer-Helfand-Hohenberg (WHH) formula[41] and saturates to $\mu_0 H_{c2}^{\perp}(0) = 2.4$ T in the ultralow temperature regime. Meanwhile, under in-plane magnetic fields, the superconductivity can survive up to $\mu_0 H_{c2}^{\parallel}(0) = 30.6$ T, which is more than 10 times larger than the Pauli limit (Chandrasekhar-Clogston limit)[42,43], *i.e.*, $\mu_0 H_p \sim g^{-\frac{1}{2}} \Delta/\mu_B \sim 3.0$ T, where $\mu_B$ is the Bohr magneton and Landé $g$-factor is assumed to be 2. The in-plane critical field is quantitively consistent with the Klemm-Luther-Beasely (KLB) theory[44] as shown by the red solid line in Fig. 1(c). Moreover, the $\mu_0 H_{c2}^{\parallel} - T_c$ relations for all samples can be well-fitted by the KLB formula, indicating that the spin-orbit scattering mechanism plays a dominant role. The estimated parameters satisfy the dirty-limit superconductors (see Table SI, Fig. S3, 6 and Supplementary Section 2, 4 for the detailed analysis). In thicker films, the spin-orbit scattering time $\tau_{so}$ becomes larger (Fig. S6(b)) as a result of the reduced $\mu_0 H_{c2}^{\parallel}$ and the weakened impurity scattering. The scattering time $\tau$ increases 100 times in the 22 nm-thick sample, which indicates the largely reduced disorder. Therefore, on the quantitative level, the influence caused by quenched disorder can be tuned by the thickness of samples. We conclude that the thickness-dependent $\beta-\mathrm{W}$ films are 2D highly-disordered and tunable dirty superconductor, which can serve as an ideal platform to explore the TUC.

We first demonstrate the critical behaviors under out-of-plane magnetic fields. To investigate the characteristics of the QPTs, the temperature-dependent magnetoresistance is analyzed in detail. As the out-of-plane magnetic field increases, the 2D superconductor gradually changes to a weakly localized metal (WLM) with a critical resistance $R_c \ll h/4e^2$. By carefully analyzing the isotherms data of the 2 nm $\beta-\mathrm{W}$ sample around the critical region in Fig. 2(a), a series of continuous varying crossing points are identified, defined as $B_c(T)$ depicted in the down-right inset of Fig. 2(a), rather than one or two crossing points as in conventional SMT [2,5,45-47]. The line of crossing points turns upwards and deviates from the mean-field WHH formula [41] below a temperature $T_m \sim 1.2$ K. The deviation is different from the linear anomaly at low temperatures induced by vortex-glass[48]. $B_c(0\ \mathrm{K})$ is two times of the fitted $B_{c2}^{MF(fit)} = 3.05$ T at zero temperature, indicating the strong quantum fluctuation at low temperatures. The pronounced quantum fluctuation, in combination with the effect of the quenched disorder[49-51], can lead to exotic critical phenomena, such as QGS[7,52], which needs to be confirmed by finite-size scaling (FSS) analysis. Around the critical regime, the resistance takes the scaling form[3,13]

$$R(B,T) = R_c f\left[(B - B_c)/T^{1/z\nu}\right], (1)$$

where $R_c$, $B_c$, $z$ and $\nu$ are the critical resistance, critical magnetic field, dynamical exponent and correlation length exponent, respectively, and $f[]$ is the scaling function with $f[0] = 1$ (see Supplementary Section 5 and Fig. S7 for details). The effective "critical" exponent $z\nu$ is summarized in Fig. 2(b), and $z\nu$ grows quickly and diverges when approaching the characteristic magnetic field $B_c^* = 5.6$ T. In stark contrast to conventional SMT or SIT[3,5], the effective "critical" exponent $z\nu$ can be well-fitted by the activated scaling law (red curve)[53]

$$z\nu \propto |B_c^* - B|^{-\nu\psi}, (2)$$

where the correlation length exponent $\nu \approx 1.2$ and tunneling critical exponent in 2D $\psi \approx 0.5$. Those evidence indicate the existence of QGS behavior associated with the infinite-randomness QCP[30,52,53]. It can be attributed to the effect of quenched disorder on the Abrikosov vortex lattice in the region of $B_{c2}^{MF} < B < B_c^*$, where rare regions of large superconducting puddles gradually emerge below the temperature $T_M \sim 1.2$ K and the exponentially small excitation energy gives rise to ultraslow dynamics with diverging effective dynamical exponent $z$ around zero temperature. Significantly, the critical exponent $z\nu$ reaches 30 at 0.05 K. We also compare the in-plane critical fields $\mu_0 H_{c2}^{\parallel}$ and critical exponents $z\nu$ in our samples with those presented in previous works that exhibit QGS phases. We find that the 2 nm-thick $\beta -$ W film exhibits an ultrahigh $z\nu$ and $H_{c2}^{\parallel}/H_p$ simultaneously, as shown in Fig. S8. Besides, the activated scaling of resistance isotherms shown in Fig. 2c (see the detail in Supplementary Section 6) further prove the QGS phase. The resistance scaling has the form[15,54,55]:

$$R = \Phi\left(\left(\frac{B-B_c}{B_c}\right) \cdot \left(ln\frac{T_0}{T}\right)^{\frac{1}{\nu\psi}}, u \cdot \left(ln\frac{T_0}{T}\right)^{-y}\right), (3)$$

where $\nu$ and $\psi$ are critical exponents, $-y$ is related to the irrelevant parameter and $ln\frac{T_0}{T}$ is the effective length scale. The estimated $\nu\psi$ is 0.6 which agrees with the QGS behavior. Moreover, the activated scaling law with an irrelevant parameter gives a good fit for the critical field in the low-temperature regime [15]: $\frac{B_c - B(T)}{B_c} \propto u \cdot \left(ln\frac{T_0}{T}\right)^{-\frac{1}{\nu\psi}-y}$.The phase diagram is summarized in Fig. 2(d).

Next, we extract the temperature-dependent magnetoresistance on thicker $\beta -$ W films and use the FSS method to analyze the critical behaviors (Fig. S10-12). The representative phase diagrams to demonstrate the QPTs are summarized in Fig. 3(a)-(c) and Fig. S13. Since all the $\beta -$W films in our measurements exhibit dirty superconductor behavior, the homogeneous superconductivity and thermal fluctuation (TF) states are separated by boundary following the WHH formula. Moreover, the entire temperature-dependent $z\nu$ of all samples by the FSS method are summarized in Fig. 3(d) (see Supplementary Section 7 and Fig. S9-12 for details).

In Fig. 3(d), in region 1 with thickness $t$ locating in $2 \leq t \leq 8$ nm, the critical exponents $z\nu$ can be extremely large near-zero temperature, and accord well with the activated scaling law of QGS phase when approaching the infinite-randomness QCP (Fig. S9). The boundary of QGS can be simulated well by the activated scaling with irrelevant corrections as shown by the red line (see the detail in Supplementary Section 6). Besides, except the regime close to zero temperature, this boundary can also be well fitted as shown by the purple dashed curves in Fig. 2(f), 3(a) and Fig. S13(a)-(b) through the theory of quantum-fluctuation-enhanced critical field proposed by Galitski and Larkin[28]: $T_c \propto T_{c0}\exp\left(-\frac{h^2}{4I}\right)$, where $h = \frac{B_c}{B_c^*} - 1$ and $I$ represent the density of disorder. The disorder level can be quantitatively evaluated by the fitting parameter $I$ which decreases from 0.059 in the 2 nm-thick samples to 0.0037 in the 6 nm-thick samples. In thicker samples, relatively lower temperatures ($T/T_{c0}$) are needed to make the system transform into the QGS universality class. The reason for this behavior is the decrease of the quantum fluctuation in thicker films, as confirmed by the shrink of the QGS region shown in Fig. 2(f), 3(a) and Fig. S13(a)-(b). In region 2 with a thickness satisfying $10 \leq t \leq 14$ nm, within the temperature regime down to 0.3 Kelvin, the QGS phase disappears and two distinctive crossing points emerge with two representative critical exponents $z\nu$. As an example, in the 10 nm-thick samples (Fig. 3(b)), the first critical magnetic field near-zero temperature is $B_c^* = 4.70$ T with the critical exponent $z\nu = 1.69 \pm 0.08$, and the second crossing point in the high-temperature regime locates at $B_x^* = 4.43$ T with $z\nu = 0.72$ (see Fig. S10 and 11). These two separate crossing points may originate from the disorder-induced superconducting puddles with a similar radius[27]. In the high-temperature regime, the magnetic field diminishes the superconducting coherence within the puddles, and the universality class manifests itself as a clean QCP described by the (2+1)D XY model[45]; while in the zero-temperature regime, the magnetic field eliminates the superconducting coherence among puddles, and quenched disorder gives rise to a dirty-limit QCP satisfying the Harris criterion[45,51]. The value is closer to the classical percolation model dominated by fermions in SMT rather than the quantum percolation model ($z\nu = 2.33$) in SIT where Cooper-pairs exhibit enough integrity dominated by boson localization[10].When the thickness further increases to $t \geq 16$ nm, the system becomes relatively more homogenous, and QCP with only one crossing point for a conventional SMT appears.

Therefore, the universality class transforms with two steps from QGS to multiple quantum criticalities and finally evolves to a single quantum criticality when thickening the film. The critical thickness of the TUC is $t_{c\perp1}{\sim}8$ nm and $t_{c\perp2}{\sim}16$ nm, respectively. To account for the TUC across three distinctive types of QPTs, we propose the following schematic scenario based on the deformation of vortex structure by quenched disorder, as shown in Fig. 3(e). In the thickest region $t \geq 16$ nm, the large mean free path $l_m$ (Table SI) indicates a low quenched disorder strength and the system is relatively homogenous with conventional single quantum criticality originated from the periodic vortex lattice structure. When the thickness is thinner to $10 \leq t \leq 14$ nm, the quenched disorder strength gradually increases as indicated by the reducing $l_m$ (Table SI), and superconducting puddles with roughly the same size present, which results in two separate crossing points corresponding to the vanishing of superconductivity within the puddle and between puddles[45]. With continuously increasing the disorder strength to the strong regime, the system becomes more inhomogeneous, and the Ohmic dissipation strength in large superconducting puddles becomes large due to the coupling to metallic channels[26], and subsequently gives rise to exponentially divergent susceptibility in these puddles[56]. This is similar to the behavior of large clusters in the random transverse field Ising model[1,7]. When the temperature is below $T_m$, the vortex liquid-like phase will freeze into a vortex glass-like phase; meanwhile, the quantum fluctuation gives rise to a pronounced enhancement of $B_c$ along the critical line[28] (Fig. 2(f) and 3(a)). When approaching zero temperature, the superconductivity puddles deform into unevenly distributed superconducting puddles with exponentially small energy excitation, namely rare regions of extremely large superconducting puddles, and the QGS universality class emerges.

Under in-plane magnetic fields, the large Zeeman splitting of electrons with opposite spin leads to Cooper pairing breaking, which can also give rise to SMT [6,9,31]. However, the SMT has not been observed in 2D superconductors under in-plane magnetic fields, not to mention the quantitatively understanding of the effect of quenched disorder on the SMT within this typical geometry. Therefore, it is indispensable to investigate the QPT in this scenario to search for the universality with different microscopic processes, although there should be universal characteristics for QPTs irrelevant to the direction of the magnetic field.

Thus, we study the critical transition region under in-plane magnetic fields in a 6 nm-thick sample. The SMT is observed with continuous crossing points $B_c$ in $R-B$ isotherms obtained and shown in Fig. 4(a) (see the original temperature-dependent magnetoresistance data in Fig. S14). The crossing points saturate at low temperatures which is distinct from the out-of-plane magnetic field scenario due to the disappearance of the vortex glass state. The red line shows the good fit of phase boundary by activated scaling method which indicates the presence of QGS phase at low temperatures. Due to low-temperature constraint ~0.3 K in our He-3 high-field facility, the extreme behavior at lower temperatures is not accessible. After performing the FSS analysis, the "effective" critical exponents $z\nu$ at varying crossing points $B_c$ are obtained, and they follow the activated scaling law $z\nu \propto |B_c^* - B|^{-0.6}$ as plotted in Fig. 4(b), indicating the existence of QGS under in-plane $B$. Moreover, the full resistance isotheorms at 0.34-0.81 K can be fitted well by the activated scaling method in Fig. 4c. The phase boundary, magnetic field-dependent $z\nu$ and resistance by activated scaling method prove the QGS phase. The $z\nu^{\parallel} = 3.6$ is similar to the out-of-plane case ($z\nu^{\perp} = 3.1$) at 0.34-0.47 K. However, the characteristic melting temperature $T_M^{\parallel}$~0.6 K is smaller than the $T_M^{\perp}$. This may originate from the different microscopic mechanisms for rare regions under in-plane $B$, as schematically shown in Fig. 4(c) inset. Although the Zeeman splitting leads to parts of coherent Cooper pairs breaking[57], large rare regions of superconducting puddles persist in the critical region near-zero temperature QCP due to quenched disorder, and the dynamics of the system becomes very slow, similar to the case under out-of-plane $B$. Furthermore, the understanding of the underlying mechanisms for SMT may be acquired by thoroughly studying the SMT features driven by varying disorder strength.

We then investigate the thickness-tuned QPTs under in-plane magnetic fields (Supplementary Section 8 and Fig. 4(d)). The thickness-dependent $z\nu$ at the lowest temperature is summarized in Fig. S18. When the sample is thinner than 8 nm, $z\nu > 1$ at low temperature. Similar to the out-of-plane magnetic field case, there exists a QGS phase in 5 and 6 nm-thick samples (Fig. S15a-c), and the QGS phase ceases to appear in samples thicker than 8 nm (Fig. S15-16). When the sample is thicker than 8 nm, $z\nu$ is smaller than 1 which corresponds to the clean critical behavior. Both the Zeeman field and flux effect play an important role in the transition region, and the FSS analysis becomes inappropriate at high temperatures (Fig. S16-17). Thus, the universality class is changed at the critical thickness of $t_{c\parallel}$~8 nm. Moreover, single crossing point-like behavior emerges in thicker samples, which is represented by the typical sample with a thickness of 22 nm, as shown in Fig. 4(d).

Consequently, the TUC behaviors are discovered under both out-of-plane and in-plane magnetic fields. The critical thickness for the transition into QGS is almost the same for out-of-plane and in-plane magnetic fields $t_{c\perp 1} = t_{c\parallel}$~8 nm, which suggests the QGS manifests itself as the final universality class under large disorder strength, regardless of the different microscopic processes. In contrast, the switch from multiple criticality to single criticality occurs at $t_{c\perp 2}$~16 nm under out-of-plane magnetic fields, while it does not exist under in-plane magnetic fields. It indicates that the superconducting puddles with roughly the same size are easier to be developed under out-of-plane than in-plane magnetic fields. This feature is unusual and needs further investigation. In retrospect, the disharmony between QGS and multiple quantum criticality is a source of great controversy in the theoretical aspect[4]. Multiple quantum criticality was previously reported in cuprate, $LaAlO_3/SrTiO_3$ interface, and superconducting arrays[45-47] and thought to be the critical state at large disorder strength in experiments. The discovery of TUC indicates that the QGS with infinite-randomness QCP universality class may act as the long-time-debated critical state under large disorder strength.

In conclusion, we have demonstrated the TUC of SMT in $\beta - \mathrm{W}$ thin films of varying thickness, with two-step switching from QGS to multiple criticality at a critical thickness of $t_{c\perp 1}$~8 nm, and finally to single criticality at $t_{c\perp 2}$~16 nm. Moreover, the SMT is observed for the first time under in-plane magnetic fields and universality of TUC. Our work not only provides an overall and self-consistent framework for the disorder effect on SMT but also opens a new frontier in the research of disorder-enriched quantum phase transitions and quantum criticality.

## Figure captions

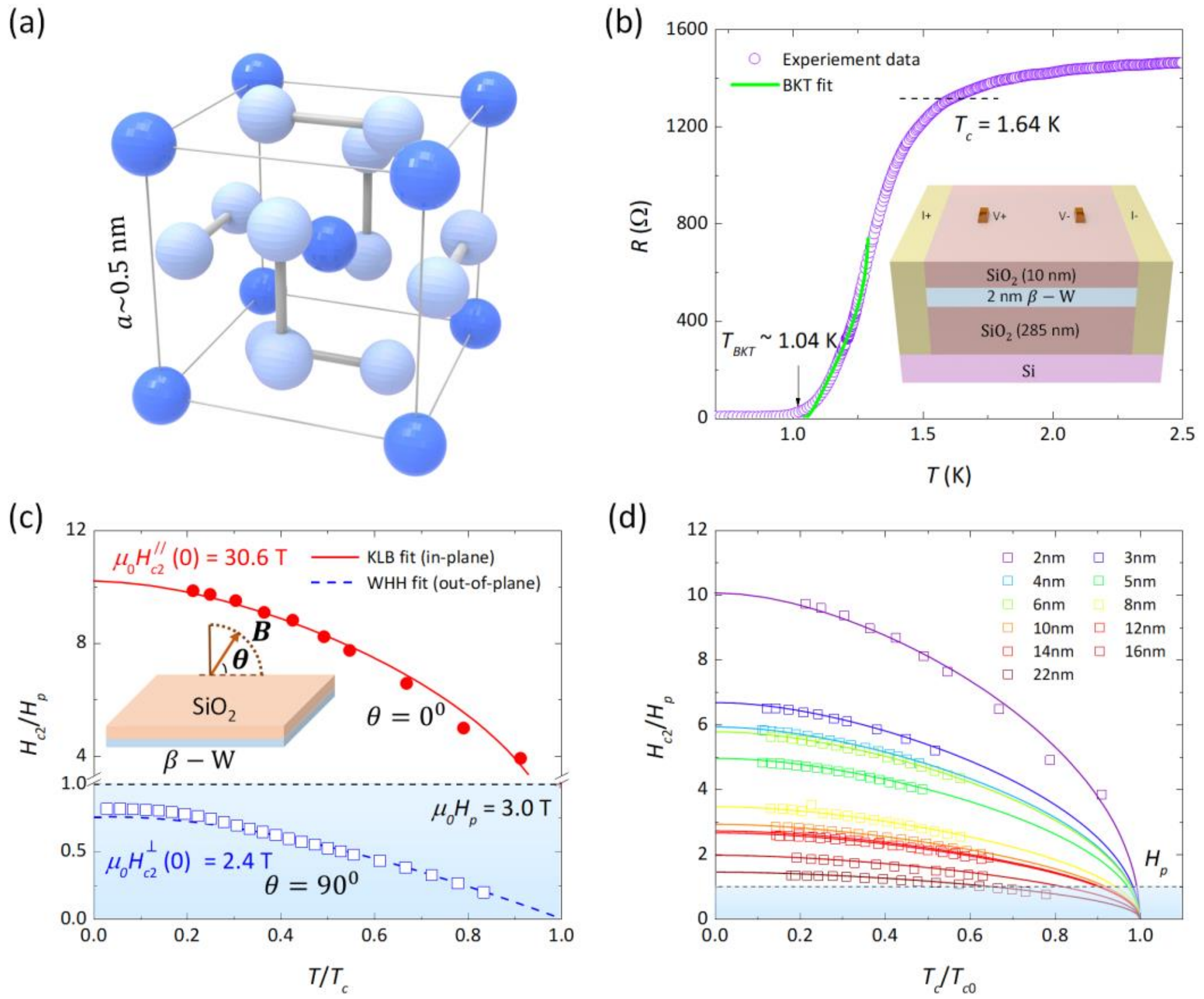

FIG. 1. 2D superconducting properties of $\beta-\mathrm{W}$ film. (a) Crystal structure of $\beta-\mathrm{W}$. Two types of W atoms in the A-15 crystal structure are shown by two colors with lattice constant $a = 5.06$Å. (b) Temperature dependence of $\beta-\mathrm{W}$ resistance for 2 nm-thick sample #01. The green solid line represents the BKT transition using the Halperin-Nelson equation[37] $R = R_0 \exp\left[-2b\left(\frac{T_{c1}-T}{T-T_{BKT}}\right)^{1/2}\right]$ with $T_{BKT} = 1.04$ K. Inset: A schematic for standard four-electrode transport measurements. (c) Temperature dependence of critical fields under perpendicular and parallel magnetic fields. The inset shows the measurement geometry. For in-plane $\mu_0H_{c2}^{\parallel}(T)$, the red line is the fitting curve using the KLB formula[44] $\ln\left(\frac{T}{T_c}\right) + \Psi\left(\frac{1}{2} + \frac{3\tau_{so}\left(\mu_0H_{c2}^{\parallel}\right)^2}{4\pi\hbar k_B T}\right) - \Psi\left(\frac{1}{2}\right) = 0$, with $T_c = 1.65$ K and $\tau_{so} = 17.6$ fs. The black dashed line corresponds to the Pauli limit field $\mu_0H_p = 3.0$ T. And, for the out-of-plane $\mu_0H_{c2}^{\perp}(T)$, the blue dashed line is the fitting curve using WHH formula[41] $\ln\left(\frac{T_0}{T}\right) = \frac{1}{2}\Psi\left(\frac{1}{2} + \frac{a+ib\mu_0H_{c2}^{\perp}}{2\pi T/T_0}\right) + \frac{1}{2}\Psi\left(\frac{1}{2} + \frac{a\mu_0H_{c2}^{\perp}-ib\mu_0H_{c2}^{\perp}}{2\pi T/T_0}\right) - \Psi\left(\frac{1}{2}\right)$, with $\Psi$ denoting the digamma function and $a = 0.335\ \mathrm{T}^{-1}$, $b = 0.2\ \mathrm{T}^{-1}$, and $T_0 = 1.8$ K. (d) Thickness-dependent in-plane upper critical fields as a function of temperature. The solid curves show the KLB fitting.

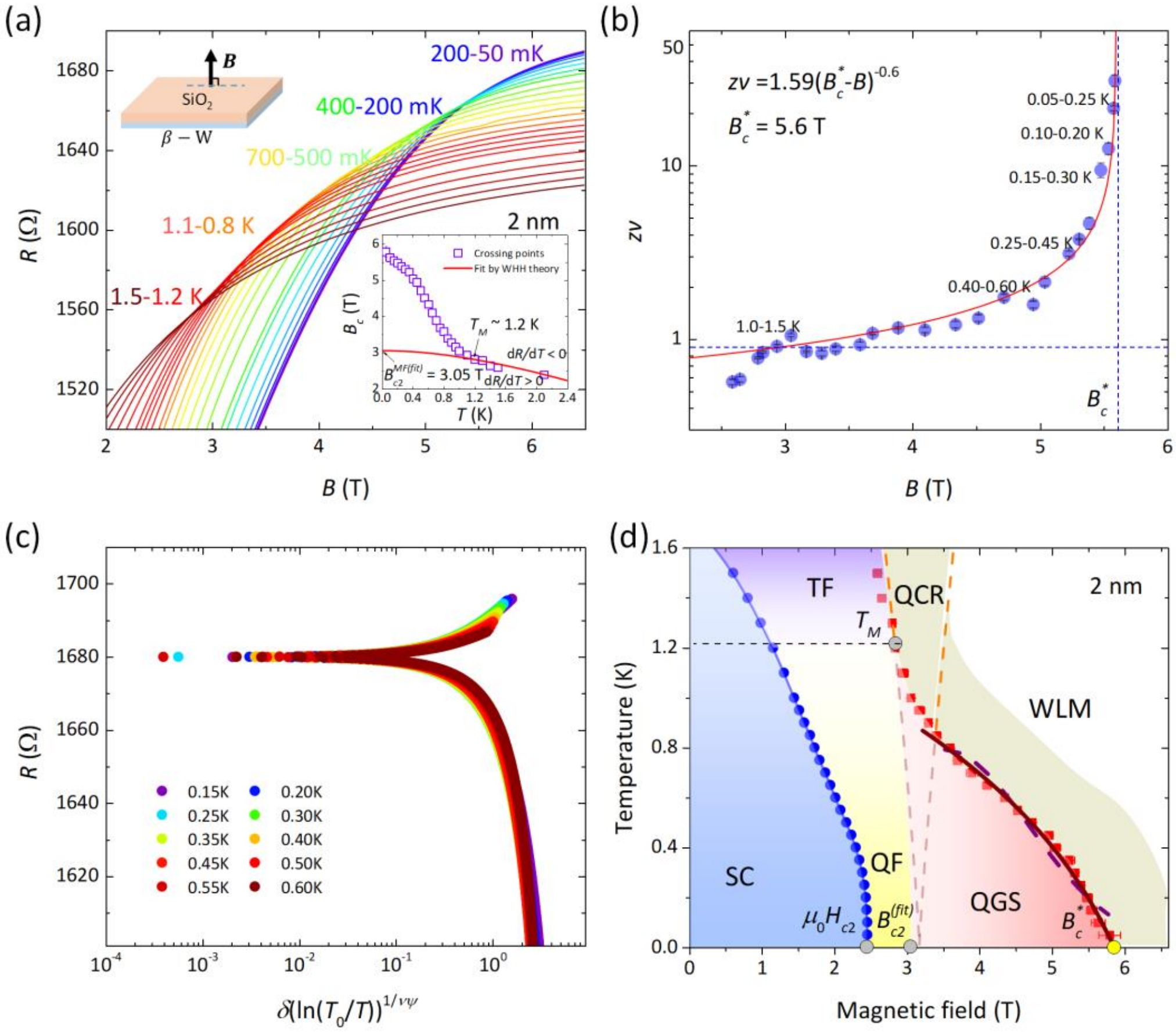

FIG. 2. The quantum Griffiths singularity behavior under out-of-plane magnetic fields (sample #01, 2 nm-thick). (a) The out-of-plane magnetic field dependence of resistance at various temperatures ranging from 0.05-1.5 K. It reveals a broad crossing region. Left inset: the experimental configuration. Right inset: the extracted crossing points $B_c$. The red line is the fitting cure using the WHH formula (1.1-2.1 K). (b) The effective critical exponent $z\nu$ (obtained by the FSS analysis in Fig. S7) as a function of $B_c$ satisfying the activated scaling law $z\nu \propto |B_c^* - B|^{-0.6}$ with $B_c^* = 5.6\ \mathrm{T}$ and indicating the QGS behavior. The purple dashed line represents the quantum Griffiths singularity region. The error bar representing the width of $z\nu$ value was acquired during the scaling analysis. (c) Activated scaling of resistance under different magnetic fields with irrelevant corrections as[15,54,55] $R = \Phi\left(\left(\frac{B-B_c}{B_c}\right)\cdot\left(ln\frac{T_0}{T}\right)^{\frac{1}{\nu\psi}}, u\cdot\left(ln\frac{T_0}{T}\right)^{-y}\right)$, where $\nu$ and $\psi$ are critical exponent, $-y$ is related to the irrelevant parameter and $ln\frac{T_0}{T}$ is the effective length scale. The fitting details are displayed in Supplementary Section 6 and the parameters are $\nu\psi = 0.6$, $y = 3$, $T_0 = 15$ K, and $B_C = 5.6$ T**.** (d) Full phase diagram for SMT quantum phase transition in 2 nm-thick $\beta - \mathrm{W}$ films. $T_{c0}$ is the transition temperature at which the resistance drops to 90% of the normal resistance, and blue dots represent the superconducting transition. The red squares with error bars show the crossing points $B_c$ of $R - B$ curves in Fig. 2(a). With the increasing magnetic field, the system transits from the superconducting state (SC) (blue region) into thermal fluctuation (TF) (purple region) or quantum fluctuation (QF) (yellow region) for high temperature and low temperature, respectively, and eventually enters the quantum Griffiths phase (red region) terminated at the infinite-randomness QCP $B_c^*$. The boundary of the QGS region can be described by the formula of Galitski-Larkin[28] (purple curve): $T_c \propto T_{c0}\exp(-h^2/4I)$, where $h = \frac{B_c}{B_c^*} - 1$ and $I$ represent the disorder strength and the activated scaling model: $\frac{B_c - B(T)}{B_c} \propto u \cdot \left(ln\frac{T_0}{T}\right)^{-\frac{1}{\nu\psi} - y}$ with parameters $B_c = 5.9$ T, $T_0 = 15$ K, $\nu\psi = 0.6$, $y = 3$, $u{\sim}60$ (red curve). The quantum critical

region (QCR) is displayed by the grey region, and WLM denotes a weak localized metal. The QGS regions shrink when increasing the film thickness.

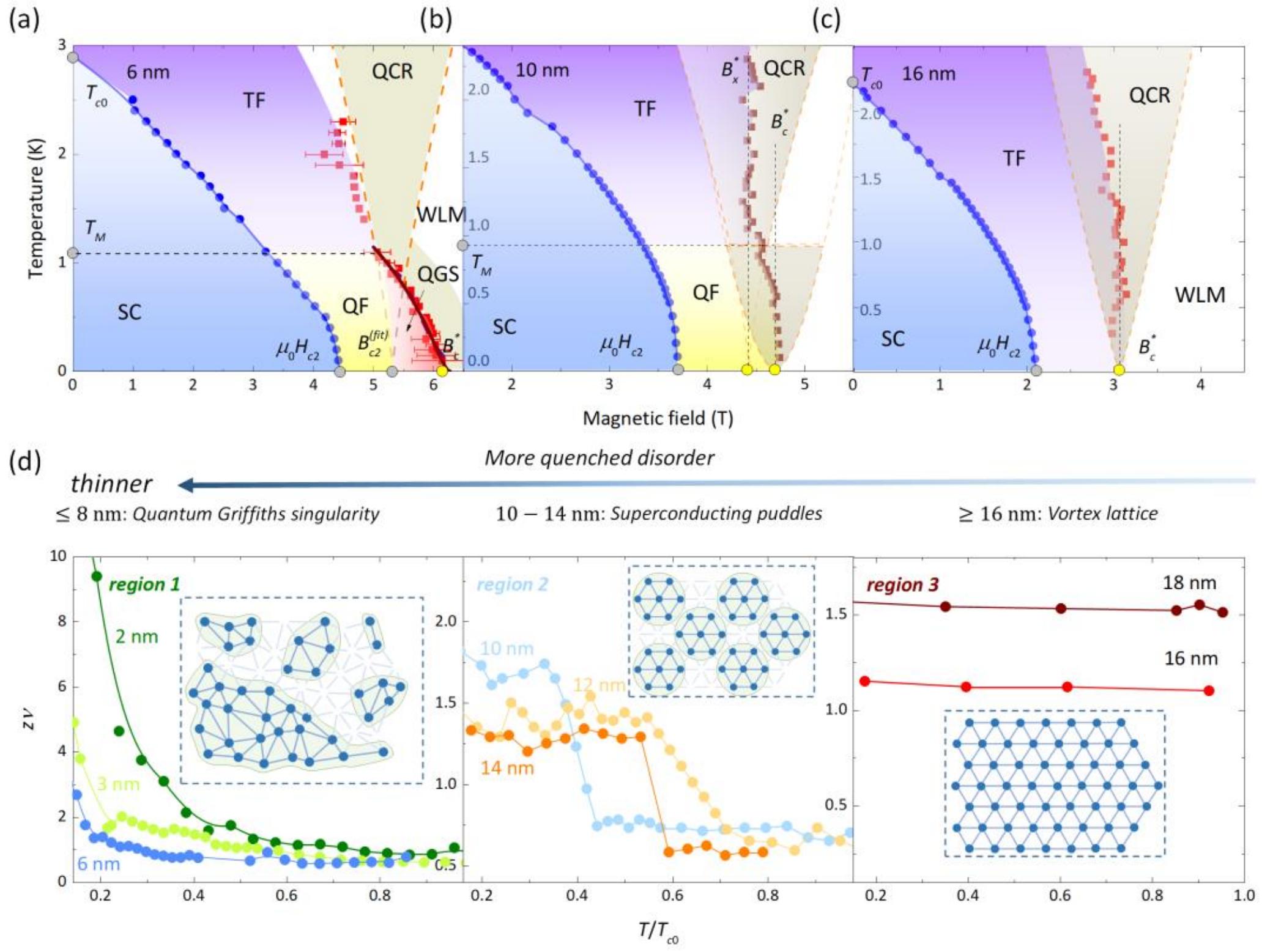

FIG. 3. The discovery of TUC by varying the sample thickness. (a-c) Full phase diagram for SMT quantum phase transition in 6, 10 and 16 nm-thick $\beta-\mathrm{W}$ films (sample #07, 10, 14), respectively. (b) In 10 nm-thick sample, two separate crossing points appear with $B_c^* = 4.70\,\mathrm{T}$ and $B_x^* = 4.43\,\mathrm{T}$, representing the low-temperature and high-temperature QCP respectively, with the associated grey critical regions (see Supplementary Section 6 and Fig. S10 and 11 for details). (c) Eventually, in 16 nm-thick samples, only one critical point with conventional scaling behavior remains(see Fig. S12 for details). (d) Normalized temperature $(T/T_{c0})$-dependent $\ln z\nu$. Three regions are divided according to the change of $z\nu$ value. Region 1: the continuous change of the $z\nu$ value satisfying QGS; region 2: two discrete $z\nu$ values; region 3: one $z\nu$ value. Inset: Illustration of disorder-induced deformation of vortex structure. In thicker films, the disorder is weak and the system behaves as a vortex lattice; in a medium thickness region (10-14 nm), the disorder is of moderate strength and superconducting puddles with nearly the same size emerge; and in films thinner than 8 nm with the large disorder, the rare regions of large superconducting puddles emerge.

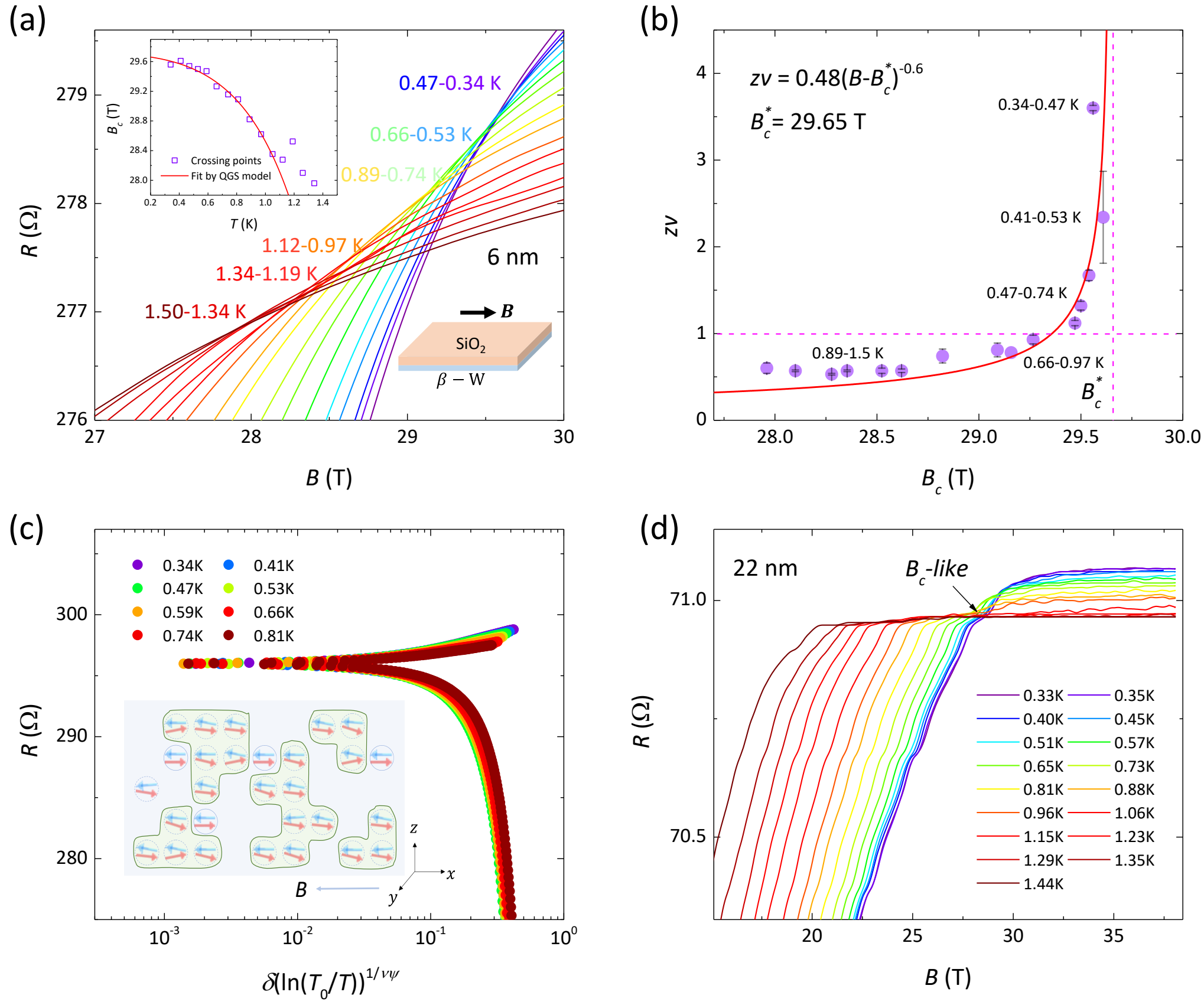

FIG. 4. SMT and TUC under in-plane magnetic fields. (a) The isotherms of resistance from 0.34-1.5 K under the in-plane magnetic field (sample #08: 3 nm-thick). Left inset: the temperature dependence of $B_c$, and the red line shows the fitting curve by 2D GL theory. Right inset: the experimental configuration. The red line is the phase boundary fit by activated scaling method with the fitting parameters $\nu \cdot \psi = 0.6$, $y = 3$, $T_0 = 5$ K, $u \sim 0.4$ and $B_C = 29.7$ T. (b) The QGS under in-plane $B$. The effective critical exponent $z\nu$ as a function of $B_c^{\parallel}$ satisfies $z\nu \propto |B_c^* - B|^{-0.6}$ with $B_c^* = 29.65$ T, as shown by the red curve. (c) The activated scaling of resistance under different magnetic fields with irrelevant corrections with the fitting parameters $\nu\psi = 0.6$, $y = 3$, $T_0 = 5$ K, and $B_C = 29.7$ T. Inset shows a schematic plot of disorder-induced deformation of Cooper pairs structure under in-plane magnetic fields. Rare regions emerge under in-plane magnetic fields and large coherent superconducting puddles are surrounded by metallic states. The large in-plane magnetic field leads to Cooper pairing breaking process due to the large Zeeman splitting of electrons with the opposite spin. (d) Resistance as a function of the magnetic field in 22 nm-thick samples (sample #16). One temperature-independent crossing point occurs and the resistance saturates at a finite magnetic field, invalidating the FSS analysis.

# Supplemental material for

## "The Discovery of Tunable Universality Class in Superconducting $\beta - W$ Thin Films"

Ce Huang[1,2†], Enze Zhang[1,2†], Yong Zhang[3†], Jinglei Zhang[4†], Faxian Xiu[1,2,5,6*], Haiwen Liu[7*], Xiaoyi Xie[1,2], Linfeng Ai[1,2], Yunkun Yang[1,2], Minhao Zhao[1,2], Junjie Qi[8], Lun Li[3], Shanshan Liu[1,2], Zihan Li[1,2], Runze Zhan[9], Ya-Qing Bie[9], Xufeng Kou[3], Shaozhi Deng[9], X. C. Xie[8,10,11]

[1] State Key Laboratory of Surface Physics and Department of Physics, Fudan University, Shanghai 200433, China

[2] Institute for Nanoelectronic Devices and Quantum Computing, Fudan University, Shanghai 200433, China

[3] School of Information Science and Technology, ShanghaiTech University, Shanghai 201210, China

[4] Anhui Province Key Laboratory of Condensed Matter Physics at Extreme Conditions, High Magnetic Field Laboratory of the Chinese, Academy of Sciences, Hefei, China

[5] Collaborative Innovation Center of Advanced Microstructures, Nanjing 210093, China

[6] Shanghai Research Center for Quantum Sciences, Shanghai 201315, China

[7] Center for Advanced Quantum Studies, Department of Physics, Beijing Normal University, 100875 Beijing, China

[8] Beijing Academy of Quantum Information Sciences, West Bld.3, No.10 Xibeiwang East Rd., Haidian District, Beijing 100193, China

[9] State Key Lab of Optoelectronic Materials and Technologies, Guangdong Province Key Laboratory of Display Material and Technology, School of Electronics and Information Technology, Sun Yat-sen University, Guangzhou, 510275, People's Republic of China.

[10] International Center for Quantum Materials, School of Physics, Peking University, 100871 Beijing, China.

[11] CAS Center for Excellence in Topological Quantum Computation, University of Chinese Academy of Sciences, Beijing 100190, China

[†]These authors contributed equally to this work.

[*]Correspondence and requests for materials should be addressed to F. X. (E-mail: Faxian@fudan.edu.cn) and H. L. (E-mail: Haiwen.liu@bnu.edu.cn)

## 1. Thickness determination of thin $\beta - \mathrm{W}$ films

All the $\beta - \mathrm{W}$ films with a thickness of 2~24 nm were prepared by a *d.c.* magneton sputtering deposition system. The system has a base pressure of $\sim 5 \times 10^{-7}$ Torr. Following a 10 min pre-sputtering, 8 W power was sequentially applied and W films were deposited onto $SiO_2$/Si substrates at an Ar sputtering pressure of $1.5 \times 10^{-3}$ Torr and a deposition rate of $1 \mathrm{~nm/min}$. Then, 10 nm-thick $SiO_2$ capping layers were grown onto $\beta - \mathrm{W}$ films with 100 W power and a deposition rate of 2 nm/min.

Four-terminal temperature-dependent transport measurements were carried out in a Physical Property Measurement System (PPMS, Quantum Design) with a dilution refrigerator (down to 35 mK) and an Oxford dilution refrigerator (KelvinoxJT) equipped with a 14 Tesla magnet (down to 45 mK). The differential resistance (d$V$/d$I$) was acquired by *a.c.*-modulation technique. Lock-in amplifiers with a low frequency (<50 Hz) was used for the transport measurements. High-magnetic-field transport experiments were carried out in water-cooled resistive magnets at the High Magnetic Field Laboratory in Hefei and the National High Magnetic Field Laboratory in Tallahassee.

To directly measure the thickness of $\beta - \mathrm{W}$ films, we performed the cross-section Transmission Electron Microscope (TEM) experiments of the W (16 min deposition) interface as shown in Fig. S1. The thickness of the W layer can be estimated to be $16 \pm 2$ nm. This helps us to confirm that the deposition rate is ~1 nm/min.

We have characterized the 2, 6 and 10 nm-thick W films with atomic force microscopy (AFM) in Fig. S2. The surface height variation is less than 0.5 nm and the particle size is less than 77 nm. We do not observe island-like single crystals. Therefore the disorder strength is mostly induced by the thickness changing rather than the surface islands or puddles. The film is uniformly disordered with the disorder of differing strengths depending upon dimensionality.

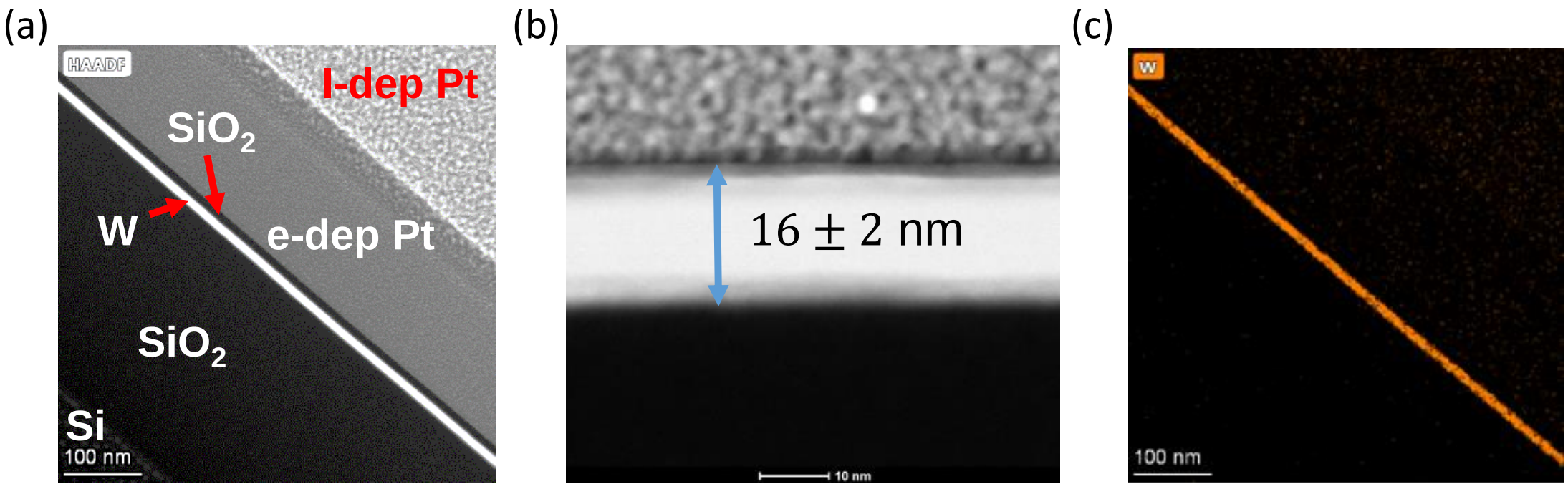

Fig. S1. TEM measurement of $\beta - \mathrm{W}$ films. (a)-(b) A cross-section high angle annular dark-field (HAADF) image of 16 nm-thick W film. (c) The EDS mapping result for W.

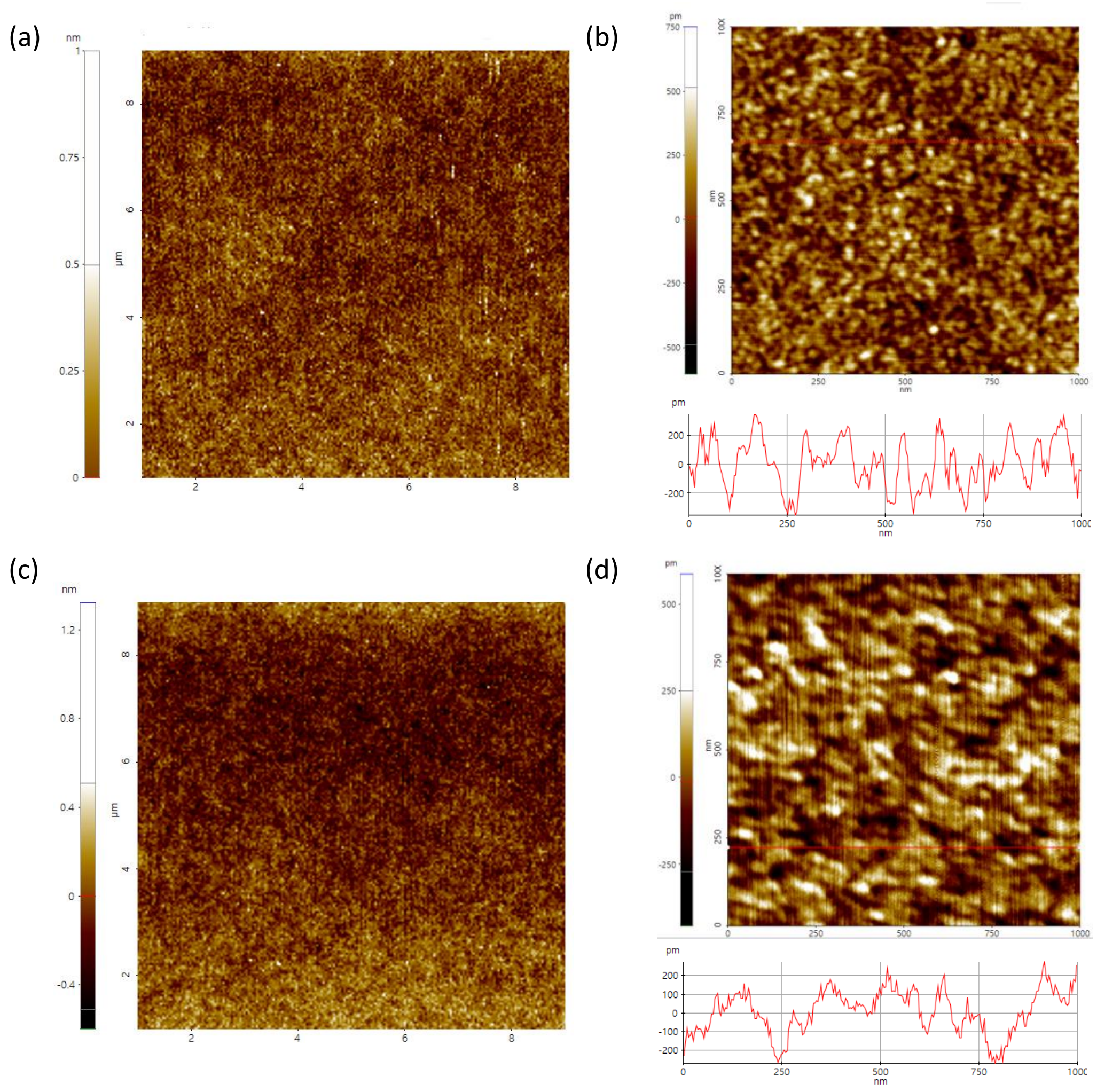

FIG. S2. AFM measurements. (a)-(b) AFM images of 2 nm-thick $\beta-\mathrm{W}$ films. (c)-(d) AFM images of 10 nm-thick $\beta-\mathrm{W}$ films. The measurements along red lines in (b) and (d) indicate the height difference smaller than 0.5 nm.

## 2. Hall effect measurement and determination of the mean free path

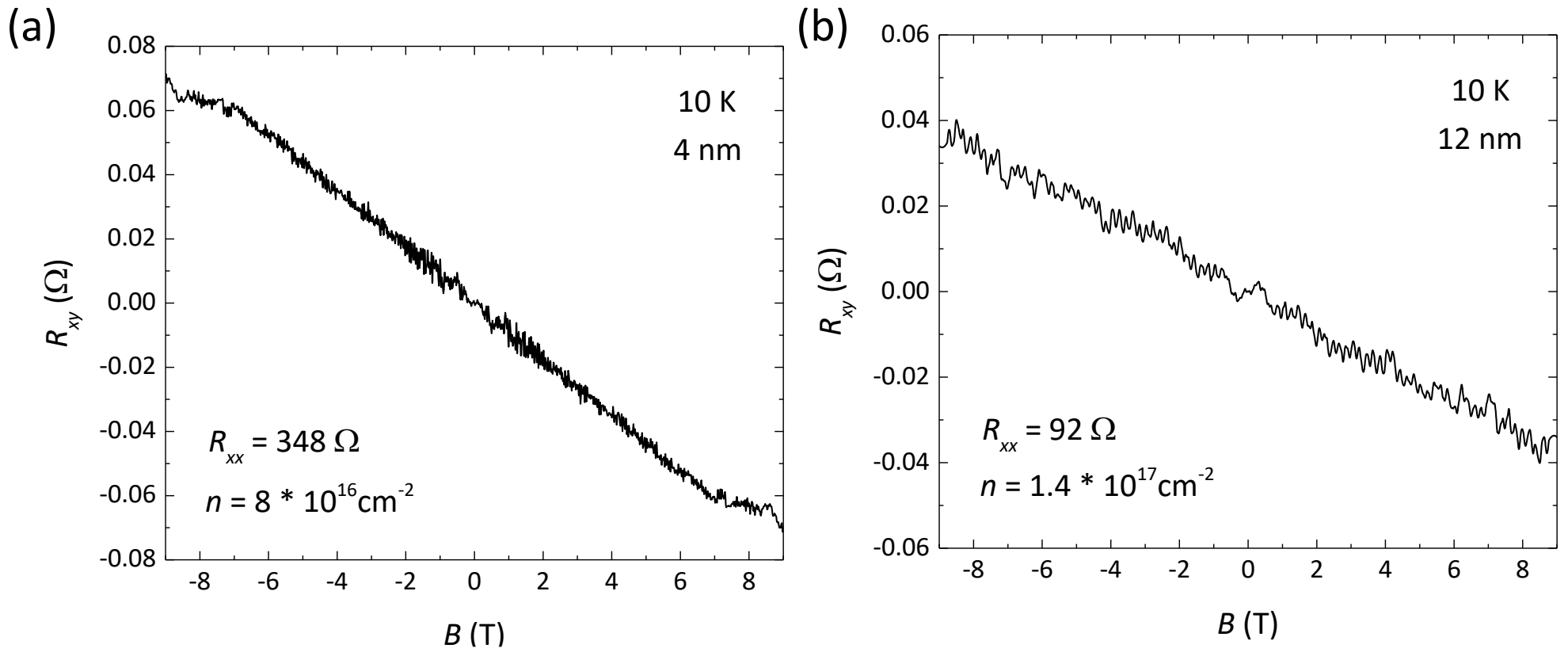

Fig. S3. Magneto-transport properties. (a)-(b) Hall resistance $R_{xy}$ of 4 nm-thick sample #05 and 12 nm-thick sample #11 at 10 K, respectively.

The magneto-transport measurements were performed at low temperatures as shown in Fig. S3. Both two samples (4 nm-thick: sample #05; 12 nm-thick: sample #11) exhibit linear Hall resistance ($R_{xy}$) which indicates one-carrier transport. The Hall mobility is quite low due to a small $R_{xy}$. The length $l$ and width $w$ of the Hall channel is $1.5\ \mathrm{cm}$ and $1.0\ \mathrm{cm}$, respectively. Since the estimation process is similar, here we use the 4 nm-thick samples for discussions. The 2D carrier density is $n_{2D} = \frac{B}{R_{xy}e} = 8 \times 10^{16}\ \mathrm{cm}^{-2}$ and then the Fermi vector is $k_F = \sqrt{2\pi n_{2D}} = 7.1\text{Å}^{-1}$. The mean free path $l_m$ is related to sheet resistance ($R_{sheet} = Rw/l$) by the formula $l_{MFP} = \frac{h/e^2}{R_{sheet}} \cdot \frac{1}{k_F} = 1.6\ \mathrm{nm}$. Besides, the resistivity of the 4 nm-thick samples is $140\ \mu\Omega \cdot \mathrm{cm}$ which accords with the previous reports[1,2] on $\beta - \mathrm{W}$ and is obviously larger than $\alpha - \mathrm{W}$ resistivity with $\rho_{\alpha-\mathrm{W}} \leq 25\ \mu\Omega \cdot \mathrm{cm}$. The transport parameters are summarized in Table SI.

## 3. 2D superconducting behavior

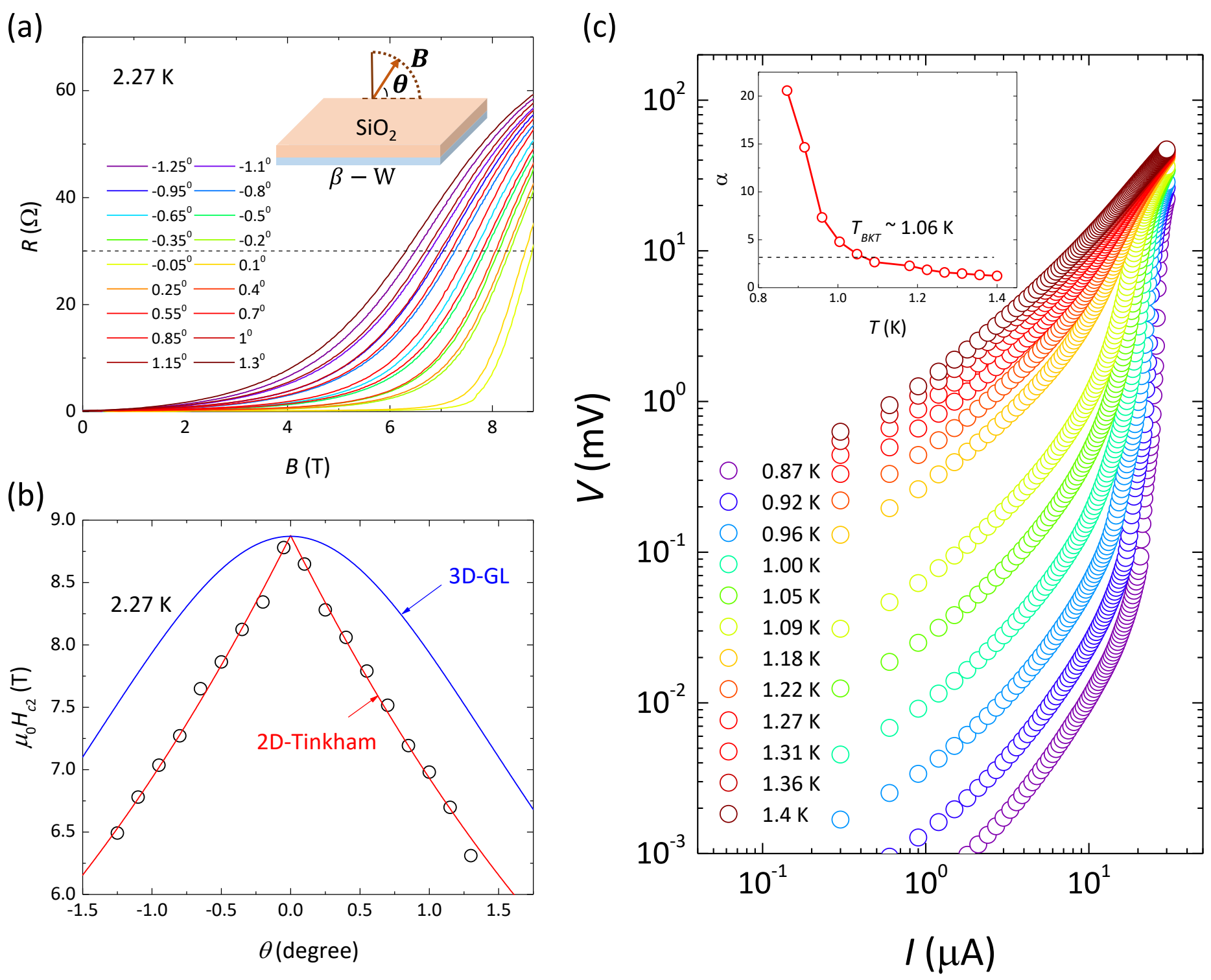

Fig. S4. 2D superconducting characteristics. (a) Angular dependence of resistance in 8 nm-thick sample #09, where the dashed line denotes $0.5R_n$. (b) Angular dependence of $\mu_0 H_{c2}(\theta)$, which is fitted well by the Tinkham formula for 2D superconductor[3] $\left((H_{c2}(\theta)\sin\theta)/H_{c2}^{\parallel}\right)^2 + |(H_{c2}(\theta)\cos\theta)/H_{c2}^{\perp}| = 1$ for a superconductor in the 2D regime ($d_{SC} < \xi_{GL}$) by the red solid line, while the experimental data largely deviate from the 3D anisotropic GL model $\left((H_{c2}(\theta)\sin\theta)/H_{c2}^{\parallel}\right)^2 + \left((H_{c2}(\theta)\sin\theta)/H_{c2}^{\perp}\right)^2 = 1$ denoted by the blue line. (c) Logarithmic scale plot of nonlinear $I-V$ characteristics in 2 nm-thick sample #01. Inset: the identification of the BKT transition. Temperature dependence of $\alpha$ exacted from the fit to the power-law dependence of $V \propto I^{\alpha}$, and the BKT transisiton temperature $T_{BKT}$~1.06 K is obtained at $\alpha = 3$.

We also performed the angle-dependent of resistance in 8 nm-thick sample and $I-V$ measurements in 2 nm-thick sample #01 as displayed in Fig. S4. The 2D Tinkham angle-dependent critical field and Berezinskii-Kosterlitz-Thouless (BKT) transition provide the evidence of the 2D superconductivity.

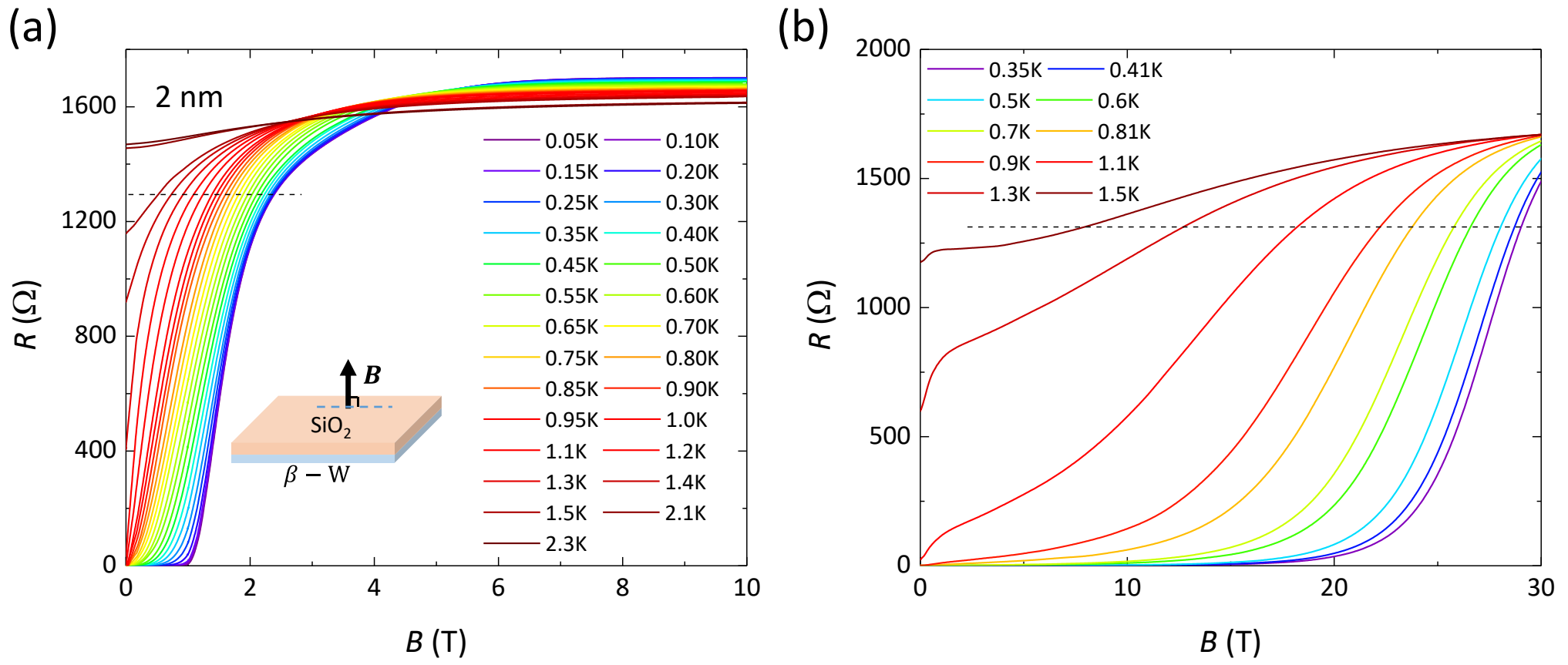

Fig. S5. Temperature-dependent magnetoresistance of 2 nm-thick $\beta-\mathrm{W}$ films as the supplementary data for Figs. 1(c), 2(a). (a)-(b) $R-B$ under out-of-plane and in-plane magnetic fields, respectively. The dashed lines show the half-normal state resistance ($0.9R_n$).

## 4. Fitting of in-plane and out-of-plane upper critical fields and related superconducting parameters

Since the thickness of sample #01 is only 2 nm which is in the 2D limit, we also try to use the 2D Ginzburg-Landau (GL) model to fit the critical fields,

$$\mu_0 H_{c2}^{\perp} = \frac{\Phi_0}{2\pi \xi_{GL}(0)^2}(1 - T/T_c)$$

$$\mu_0 H_{c2}^{\parallel} = \frac{\sqrt{12}\Phi_0}{2\pi \xi_{GL}(0) d_{SC}}\sqrt{(1 - T/T_c)}$$

where $\Phi_0$, $\xi_{GL}(0)$ and $d_{SC}$ denote the quantum flux, the GL coherence length at zero temperature, and the effective SC thickness, respectively. The results are displayed in Fig. S6(a) (red curve). $d_{SC}$ is much smaller than $\xi_{GL}(0)$ which evidences the 2D superconductivity.

We note that the 2D phenomenological GL fit for $\mu_0 H_{c2}^{\parallel}$ shows a slight deviation at low temperatures, because the phenomenological theory works near $T_c$ and cannot incorporate the microscopic spin-orbit scattering. The enhancement of $\mu_0 H_{c2}^{\parallel}$ in a dirty-limit superconductor with strong spin-orbit interaction (SOI) can be attributed to the randomization of electron spins, which results in the suppression of spin paramagnetism[4,5]. We use the Klemm-Luther-Beasely (KLB) theory[5] to fit the data

as described by

$$\ln\left(\frac{T}{T_c}\right)+\Psi\left(\frac{1}{2}+\frac{3\tau_{so}\left(\mu_0 H_{c2}^{\parallel}\right)^2}{4\pi\hbar k_B T}\right)-\Psi\left(\frac{1}{2}\right)=0,$$

where $\tau_{so}$ is the spin-orbit scattering time. Fig. S6b shows good fits by the KLB theory in 6 nm-thick sample #08. The strong SOI has been found in $\beta-\mathrm{W}$ and used to fabricate the spin Hall devices[1,2]. The KLB theory can fit well in all samples as shown in Fig. 1(d) and Fig. S6(a). Moreover, we take the 2 nm-thick samples as an example to demonstrate the related superconducting parameters. To be more specific, in the 2 nm sample, the mean free path $l_m = 0.5$ nm, $\xi_{GL} = 12$ nm, and $\tau_{so} = 18$ fs. We assume the velocity $v_F = 4.7\times10^5$ m/s, and the mean free time is estimated to $\tau = l_m/v_F = 0.1$ fs, which is typically smaller than the spin-orbit scattering time. Due to the very small mean free path with $l_m \ll \xi_{GL}$, the superconducting $\beta-\mathrm{W}$ samples are in the dirty-limit. The whole data for all samples are summarized in Table SI.

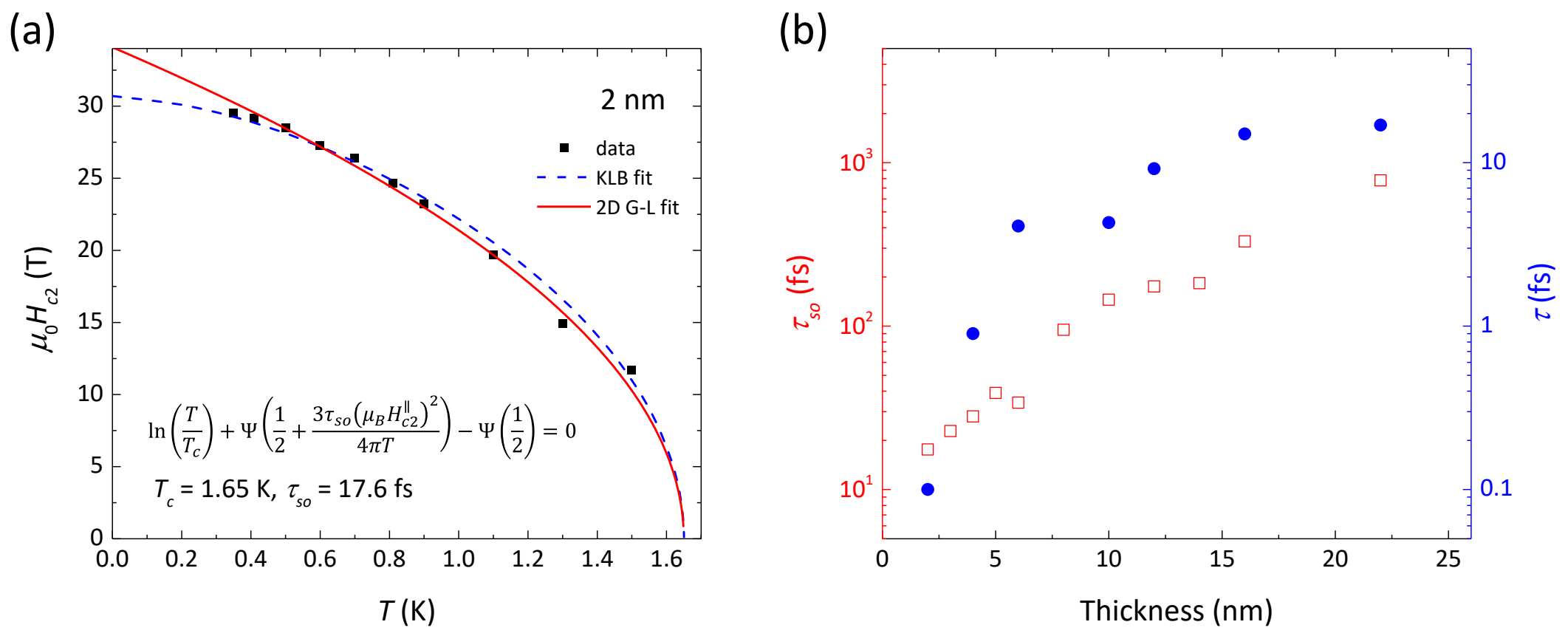

Fig. S6. The discussions on high $\mu_0 H_{c2}^{\parallel}$ by the KLB and Ising SC models. (a) Temperature-dependent in-plane critical field $\mu_0 H_{c2}^{\parallel}$ in 2 nm-thick sample #01. The blue dashed and red solid show the fitting by KLB and 2D GL, respectively. (b) The thickness-dependent $\tau_{so}$ and $\tau$ plotted on a logarithmic scale. The data are extracted from Fig. 1d and Table SI. Left red axis with red hollow squares and right blue axis with solid dots correspond to $\tau_{so}$ and $\tau$, respectively.

## 5. Finite-size scaling analysis of quantum Griffiths singularity

The small critical region formed by three adjacent curves can be treated approximately as one "critical" point $(R_c, B_c)$. We define $R(B-B_c, T_0) = R_c f[B-B_c]$ for the lowest temperature $T_0$ of the group. Then we plot the scaling curves of $R/R_c$ against the scaling variable $t|B-B_c|$. $t$ at each temperature is determined by performing a rescale of $t|B-B_c|$ to make $R/R_c$ at $T$ match the curve at the lowest temperature $T_0$. The effective critical exponent $z\nu$ is acquired by the linear fitting between $\ln(T/T_0)$ and $\ln t$ from the definition $t = (T/T_0)^{-1/z\nu}$.

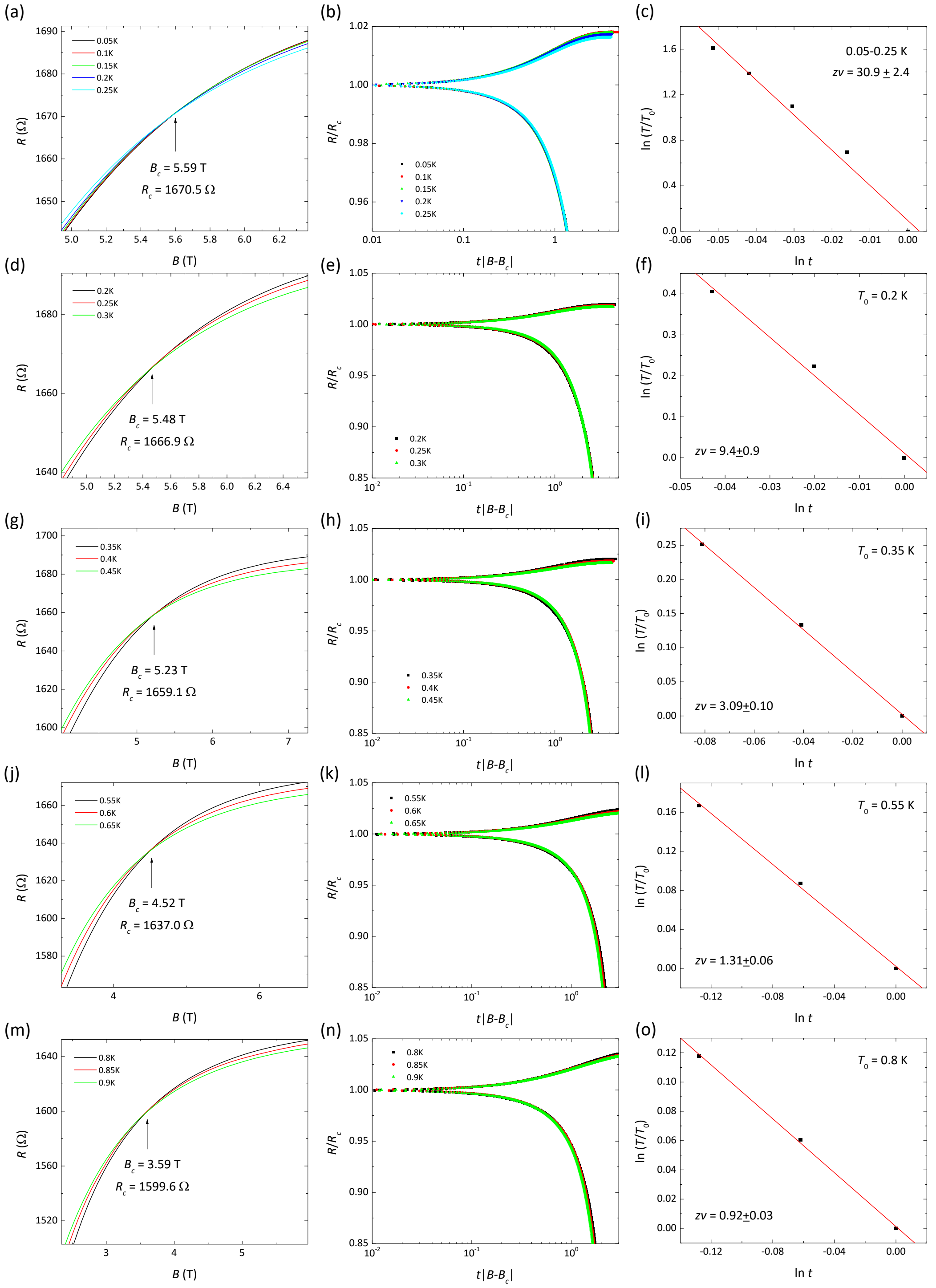

Fig. S7. FSS analysis at different temperatures along the out-of-plane magnetic field for 2 nm-thick sample #01. (a)(d)(g)(j)(m) Resistance as a function of magnetic field close to SMT boundary in the various temperature ranges of (a) 0.05-0.25 K, (d) 0.2-0.3 K, (g) 0.35-0.45 K, (j) 0.55-0.65 K and (m) 0.8-0.9 K. (b)(e)(h)(k)(n) Corresponding normalized resistance as a function of scaling variable $t|B-B_c|$, with $t = T/T_0{}^{-1/z\nu}$. (c)(f)(i)(l)(o) Corresponding linear fitting between $\ln(T/T_0)$ and $\ln t$ gives the critical exponent $z\nu$.

Especially, the curves at the lowest temperature in the 2 nm-thick sample (out-of-plane) are very close to each other as shown in Fig. S7(a). To reduce accidental errors and achieve a reliable fitting, we use five curves to perform the FSS analysis. In contrast, the scaling phenomena are easy to distinguish at all temperatures from the curves of the 6 nm-thick sample under in-plane $B$. Therefore, we use three adjacent curves to analyze the QGS.

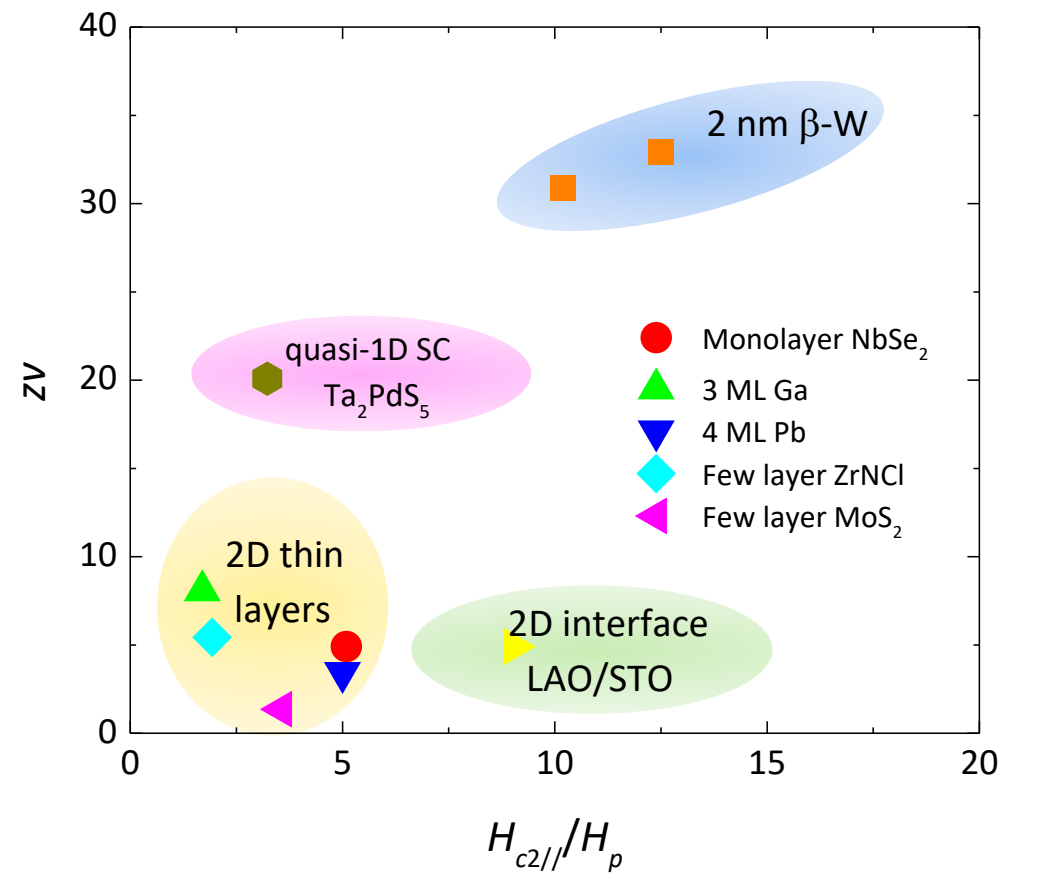

Fig. S8. Comparison of the $z\nu$ value and $H_{c2\parallel}/H_p$ among various superconducting QGS systems. The thickness of two samples (#01 and 02) are both 2 nm. The data shown are obtained from previously reported systems, including 2D Ising superconductors[6-9], thin layers metal substance[10-13], 2D interface[14-16] and quasi-1D superconductor[17].

## 6. Activated scaling for the QGS phase

We carefully analyze the data of our work based on the activated scaling analysis associated with the QGS. The strategy is to follow the procedure considering the irrelevant scaling correction on the activated scaling, provided by N. A. Lewellyn, et al (Eq. (A11) in the Appendix of [18]). The numerical scaling procedure including the irrelevant scaling variable is based on the procedure given by Slevin and Ohtsuki (Eqs. (1)-(4) in [19]). We consider the activated scaling with irrelevant corrections as follows [18]:

$$R = \Phi\left(\left(\frac{B-B_c}{B_c}\right)\cdot\left(ln\frac{T_0}{T}\right)^{\frac{1}{\nu\psi}}, u\cdot\left(ln\frac{T_0}{T}\right)^{-y}\right), \text{(S1)}$$

here $\nu$ and $\psi$ are critical exponents and $-y$ is related to the irrelevant parameter. Both the relevant and irrelevant parameters are associated with the activated scaling form with effective length scale $ln\frac{T_0}{T}$. In the numerical fitting procedure, we consider the relevance correction to the third order and the irrelevant correction to the first-order [19]. The irrelevant parameter also changes the phase boundary of the system (Eq. (A17) in the Appendix of [18]):

$$\frac{B_c-B(T)}{B_c} \propto u\cdot\left(ln\frac{T_0}{T}\right)^{-\frac{1}{\nu\psi}-y}. \text{(S2)}$$

We note the factor $-y$, which is related to the irrelevant parameter, is important to fit our experimental data. The influence of irrelevant parameter is pronounced in related experiments, but the irrelevant parameter is ignored in the previous theoretical consideration[20] [Eq. (23) in[20]]. We utilize Eqs. (S1)-(S2) to analyze the resistivity data and fitting the phase boundary. The numerical nonlinear fitting procedure based on Eq. (S1) gives the activated scaling parameters summarized as follows.

**Table S1 | Fitting parameters by activated scaling method**

| Sample | $B_c$ (T) | $T_0$ (K) | $\nu \cdot \psi$ | $y$ | $R_c$ ($\Omega$) | Fit temperature region |
|---|---|---|---|---|---|---|
| #01 | 5.6 | 15 | 0.6 | 3 | 1680 | $0.15\ \mathrm{K} \leq T \leq 0.55\ \mathrm{K}$ |
| #07 | 29.7 | 5 | 0.6 | 3 | 296 | $0.34\ \mathrm{K} \leq T \leq 0.81\ \mathrm{K}$ |

## 7. FSS analysis in samples with different thickness and supplementary data under out-of-plane magnetic fields

The QGS analysis for 3, 4 and 6 nm-thick samples are shown in Fig. S9. After the FSS analysis, similar to the 2 nm sample (Fig. 2(a)-(b)), we observe the divergent trend of $z\nu$ when approaching the infinite-randomness quantum critical point $B_c^* = 8.32$ T, which follows the activated scaling law.

The FSS analysis for 10 and 16 nm-thick samples are shown in Fig. S10 and Fig. S12(b)-(c), respectively. Moreover, to confirm the two crossing points and one crossing point in 10 and 16 nm-thick samples, we have performed the detailed temperature-dependent resistance measurements in Fig. S11 and Fig. S12(a). The temperature-independent resistance region is indicated by the critical line, which is the signature of the QPT. The two resistance plateaus in Fig. S11 reveal that two critical regions exist which correspond to the two crossing points in Fig. S10. The magnetic field and temperature value at each crossing point are close to the crossing point by $R-B$ measurements in Fig. S10(a) and (d). On the contrary, only one plateau at a wide temperature region is observed in the 16 nm-thick sample which indicates one crossing point. We note that the fit of $\ln(T/T_0) \sim \ln T$ deviates at 0.2 K in Fig. S10(c) and Fig. S12(c), which indicates that the system may start entering the QGS phase.

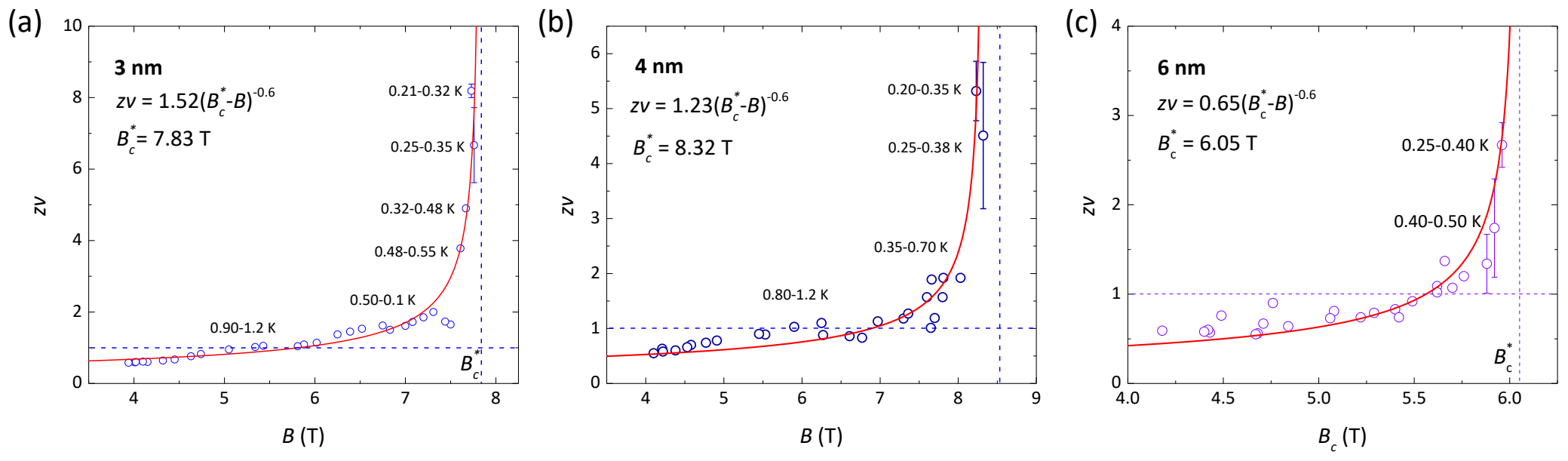

Fig. S9. QGS in 3 nm, 4 nm and 6 nm-thick samples (#03, 04 and 07, respectively). The activated quantum scaling behavior: $z\nu$ as a function of crossing points $B_c$. The red curve is fitted by $z\nu \propto |B_c^* - B|^{-0.6}$. The error bar representing the width of $z\nu$ value in (a)-(c) was acquired during the scaling analysis.

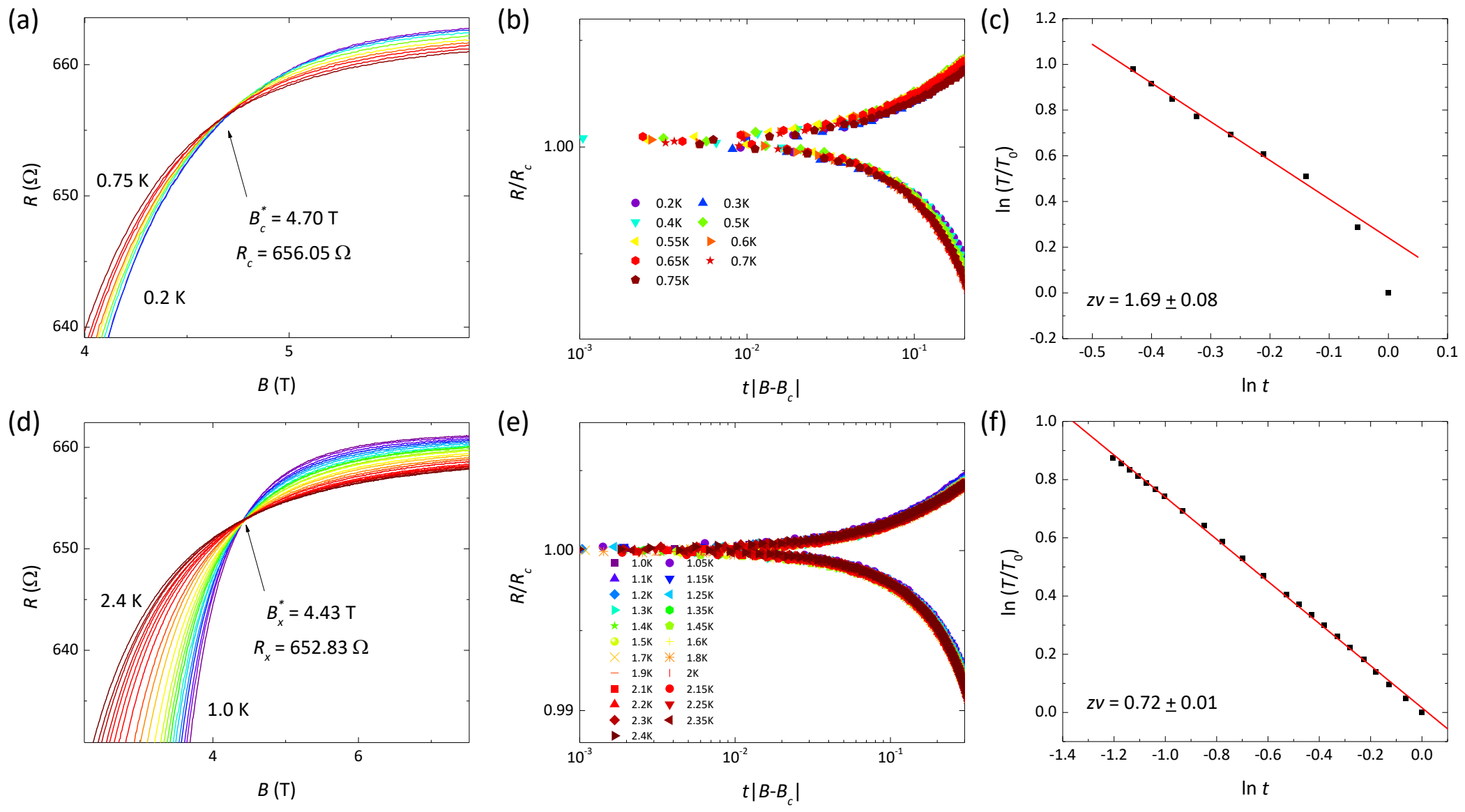

Fig. S10. FSS analysis for the 10 nm-thick sample (#10). (a) Resistance as a function of $B$ for different temperatures from 0.2 to 0.75 K. The crossing point is at $B_c^* = 4.70$ T, $R_c = 656.05\ \Omega$. (b) FSS plot of $R/R_c$ as a function of scaling variable $t|B - B_c|$, with $t = T/T_0{}^{-1/z\nu}$. (c) Corresponding linear fitting between $\ln(T/T_0)$ and $\ln t$ gives the critical exponent $z\nu = 1.69 \pm 0.08$. (d)-(f) Resistance as a function of $B$, FSS plot of $R/R_x$ and the corresponding linear fitting between $\ln(T/T_0)$ and $\ln t$ for different temperatures from 1.3 to 2.2 K, respectively. The crossing point is at $B_x^* = 4.43$ T, $R_x = 652.83\ \Omega$. The critical exponent is $z\nu = 0.72 \pm 0.01$.

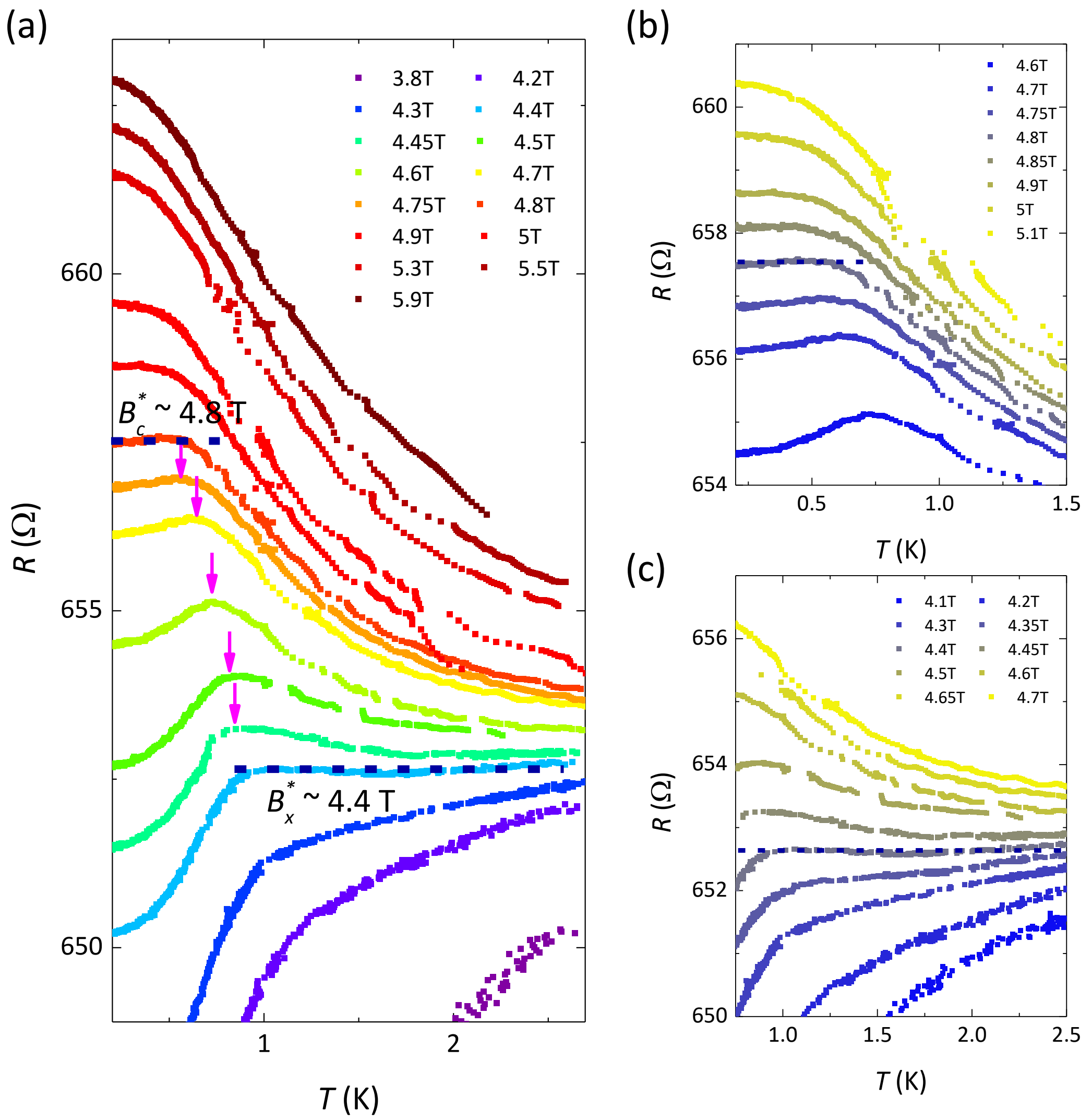

Fig. S11. Two QPTs observed by magnetic field-dependent *R*-*T* measurements in the 10 nm-thick samples (#10). (a) The resistance *R* as a function of temperature *T* for different magnetic fields from 3.8 to 5.9 T. The pink arrows indicate the *R* peaks. (b-c) The detailed data collected from the low-temperature region (0.2-1.5 K) and high-temperature region (1.0-2.5 K) with the magnetic fields varying from 4.6 to 5.1 T and from 4.1 to 4.7 T, respectively. The dashed lines indicate the critical fields $B_x^{*}$~4.4 T and $B_c^{*}$~4.8 T, respectively, where the *R* values are independent of the temperature.

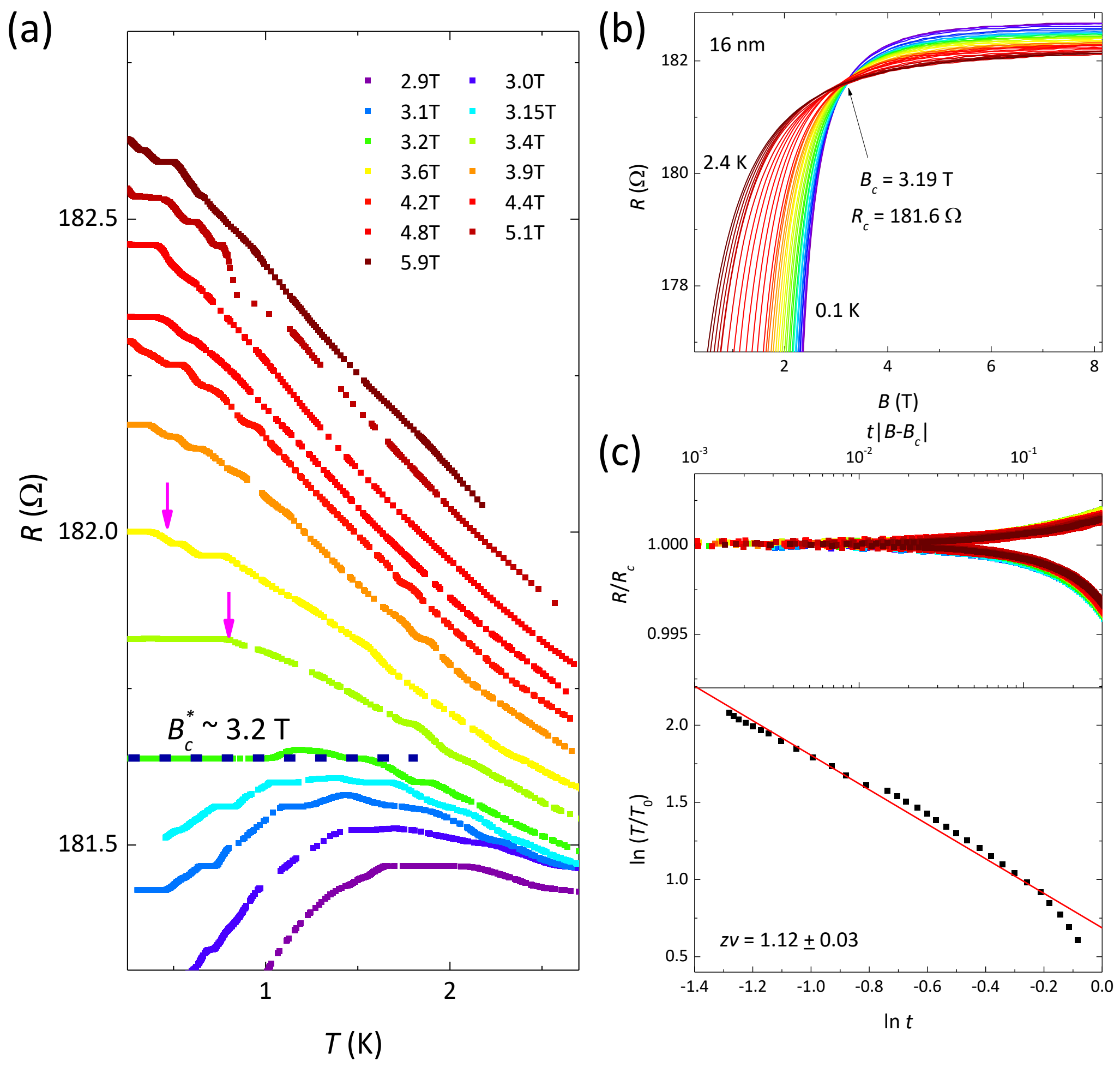

Fig. S12. One critical point in the 16 nm-thick sample (#14). (a) The resistance $R$ as a function of temperature $T$ for different magnetic fields from 2.9 to 5.9 T. The pink arrows indicate the $R$ peaks. (b) Resistance as a function of $B$ for different temperatures. The crossing point is at $B_c = 3.19$ T, $R_c = 181.6\ \Omega$. (c) FSS plot of $R/R_c$ as a function of scaling variable $t|B - B_c|$, with $t = T/T_0{}^{-1/z\nu}$. Inset: Corresponding linear fitting between $\ln(T/T_0)$ and $\ln t$ gives the critical exponent $z\nu = 1.12 \pm 0.03$.

The rest phase diagram data for 3, 4, 14 and 18 nm-thick samples are presented in Fig. S13. Both 3 and 4 nm-thick samples exhibit the QGS phase at low temperatures. The disorder density parameter is fitted to be $I = 0.030$ and $0.024$ for 3 and 4 nm thick samples, respectively (purple curves). 14 nm-thick sample exhibits two quantum critical points, which is similar to the 10 nm-thick samples. Lastly, an 18 nm-thick sample has only one crossing point which is classified as the same region with the 16 nm sample. All the phase diagrams (Fig. 2(f), Fig. 3(a)-(c) and Fig. S13) build up the evolution of thickness-dependent QPTs.

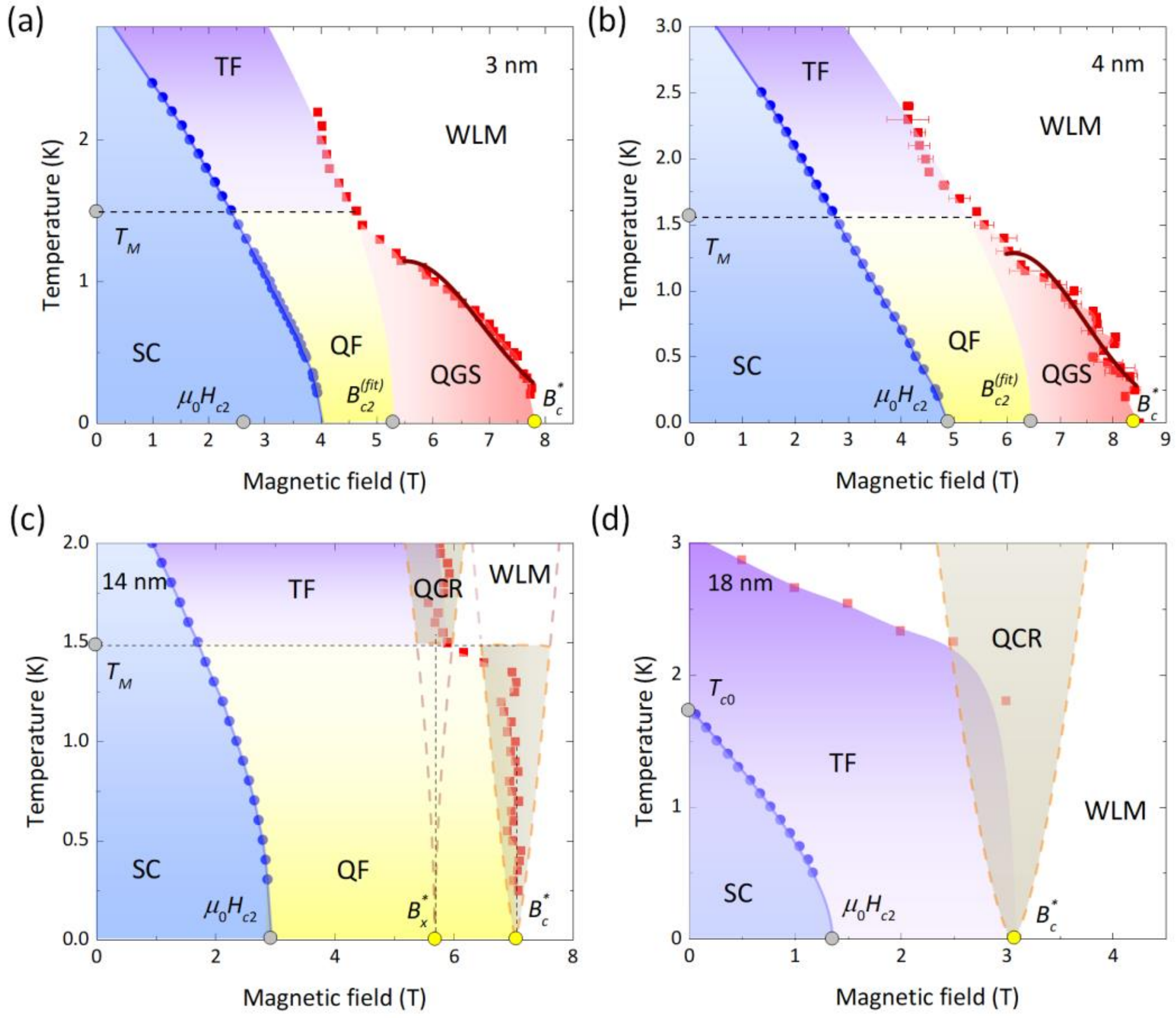

Fig. S13. Supplementary phase diagrams in the main text. Evolution of thickness-dependent phase diagram with quantum criticality under out-of-plane magnetic fields. (a)-(d) Full temperature-magnetic field phase diagrams for superconducting to metal quantum phase transition in 3 nm, 4 nm, 14 nm and 18 nm-thick $\beta-\mathrm{W}$ films (samples #03, 04, 12, 15), respectively. The details of fitting curves, plots, and phase regions are similar to Fig. 3 and can be found in the figure caption of Fig. 3. The boundary of the QGS region can be described by the formula of Galitski-Larkin[21] (purple curves): $T_c \propto T_{c0}\exp(-h^2/4I)$, where $h = \frac{B_c}{B_c^*} - 1$ and $I$ represent disorder strength.

## 8. Thickness-dependent phase diagrams and FSS analysis under in-plane magnetic fields

The $R$-$B$ measurements under in-plane $B$ and detailed FSS analysis are conducted. The critical exponent $z\nu$ for samples with a thickness of 5 nm and 6 nm are shown in Fig. S15(b), and the QGS in the former sample is stronger than the latter as evidenced by the larger $z\nu$ value at 0.34 K. Moreover, in the 5 nm-thick sample, below $T_m(5\text{ nm}) \sim 0.8$ K, $z\nu$ is greater than 1 indicating the QGS phase, as shown in Fig. S15(b). A representative phase diagram of the 6 nm-thick sample under in-plane $B$ is shown in Fig. S15(c), with the QGS region much smaller than the case under out-of-plane $B$ (Fig. 3(a)). This difference may originate from a smaller quantum fluctuation effect of the Zeeman-field Cooper pairing breaking scenario under in-plane $B$ compared with the vortex glass-like scenario under out-of-plane $B$, since for the latter case the depinning and deformation of vortex configuration provide more route to fluctuation. The Cooper pairs in the rare regions are easier to be broken by $B_{\parallel}$ than the scenario in $B_{\perp}$.

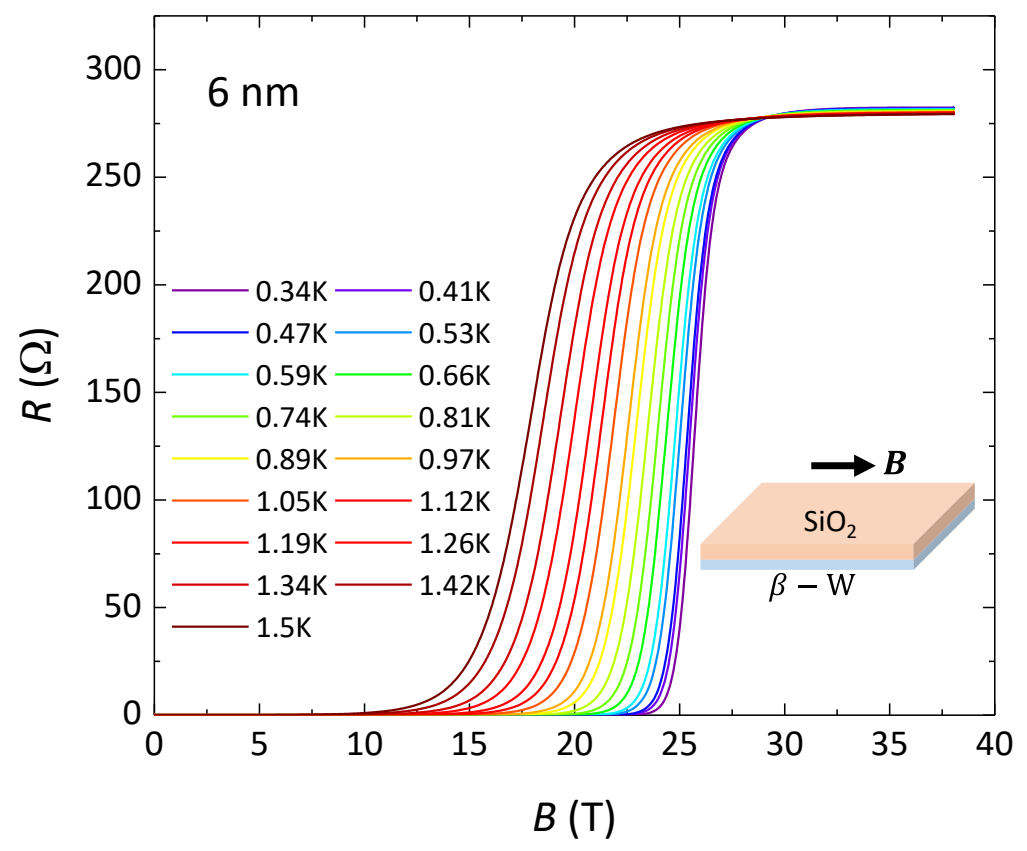

Fig. S14. Temperature-dependent magnetoresistance of 6 nm-thick $\beta - \mathrm{W}$ films as the supplementary data for Fig. 4(a). The dashed lines show the half-normal state resistance ($0.9R_n$).

Ulteriorly, when the sample thickness is larger than 8 nm at high temperature, both the Zeeman field and flux effect play an important role in the transition region, and the FSS analysis becomes inappropriate, as shown in Fig. S16-17. Firstly, the resistance starts to saturate rather than continuous increases in 5 and 6 nm-thick samples with SMT. Secondly, when the magnetic field is close to the crossing points region, the resistance shows oscillation-like behavior. The magnetic length at crossing points region 25 T is $l_B = \sqrt{\hbar/eB} \approx 5\ \mathrm{nm}$ which approaches the sample thickness. It indicates that the orbital effect is induced. The SMT is thus no longer a single variable controlled phase transition which is very complicated. We suggest that this may contribute to the oscillation behavior.

After FSS analysis, we found that the SMT is indeed complicated as shown in Supplementary Figs. 16-17 for the 8 nm-thick sample. In the 8 nm-thick sample, the $R - B$ curves are close to each other at low temperatures, and FSS seems to be applicable in the low-temperature regime, as shown in Fig. S16(b). However, FSS is not fitted well at other temperatures as shown in Fig. S16(d) and (f) ($0.59 - 0.74$ K and $1.34 - 1.5$ K, respectively). Moreover, in the thickest sample (22 nm), one QCP-like behavior is observed in Fig. 4(d). However, this QCP cannot be analyzed by the FSS theory (Fig. S17(a)-(b)) and the resistance saturates to the same value at high temperatures (Fig. S17(c)). To sum up, when the thickness is larger than 8 nm, the magnetic length at high magnetic field $l_B = \sqrt{\hbar/eB}$ will be close and even smaller than the sample thickness. The SMT is very complicated and deserves future research.

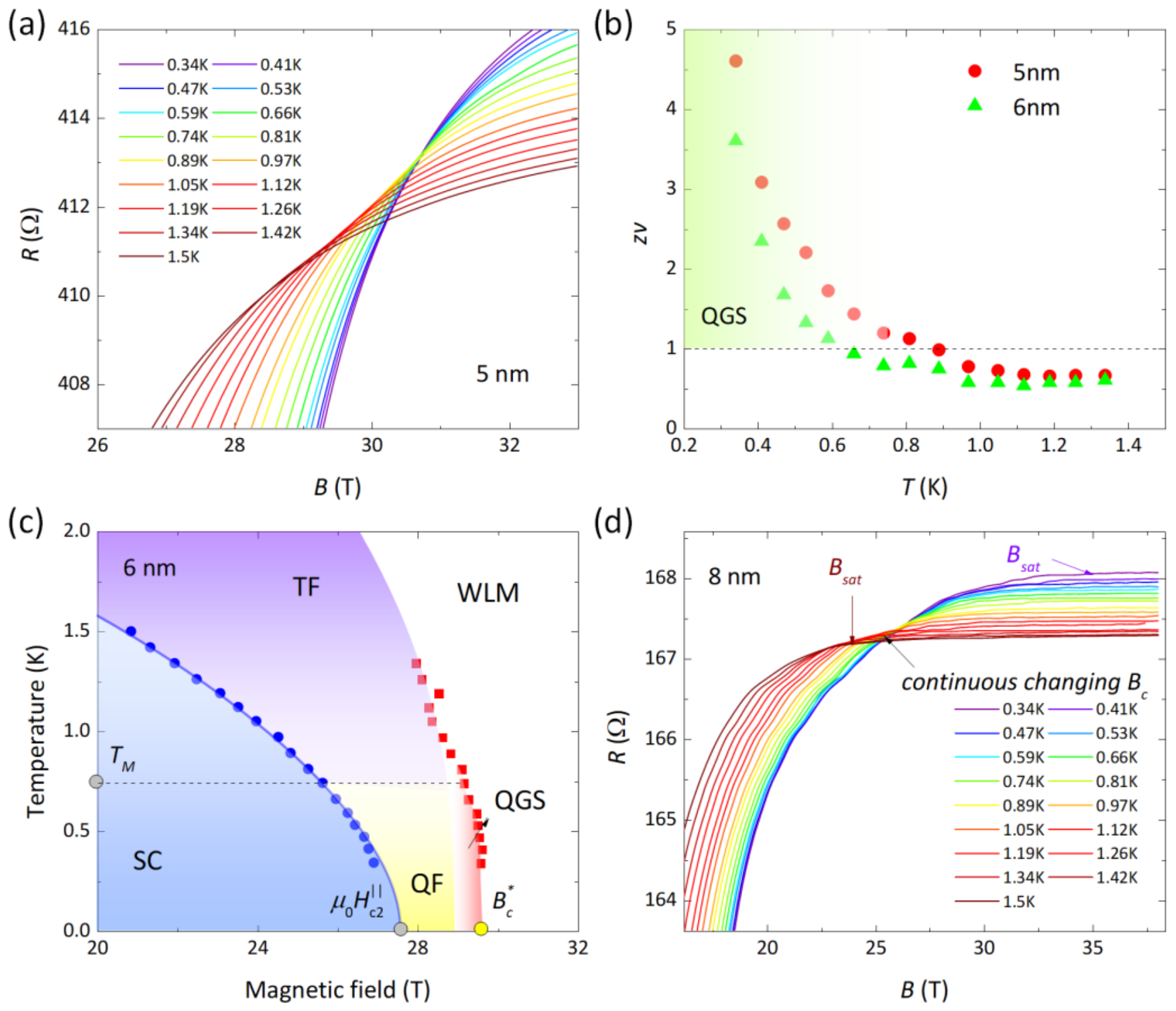

Fig. S15. The magnetic field dependence of resistance at various temperatures ranging from 0.34 K to 1.5 K. (a) and (d) $R-B$ for 5 nm and 8 nm (#06 and 09), respectively. (b) Temperature-dependent $z\nu$ in 5 nm-thick (sample #06) and 6 nm-thick samples (sample #08) by red and green dots, respectively. (c) The phase diagram in 6 nm $\beta-\mathrm{W}$ under in-plane magnetic fields. At high magnetic fields and low temperatures, the QGS phase occurs close to quantum critical field $B_c^*$. The blue line is the fitting curve by the KLB formula[5].

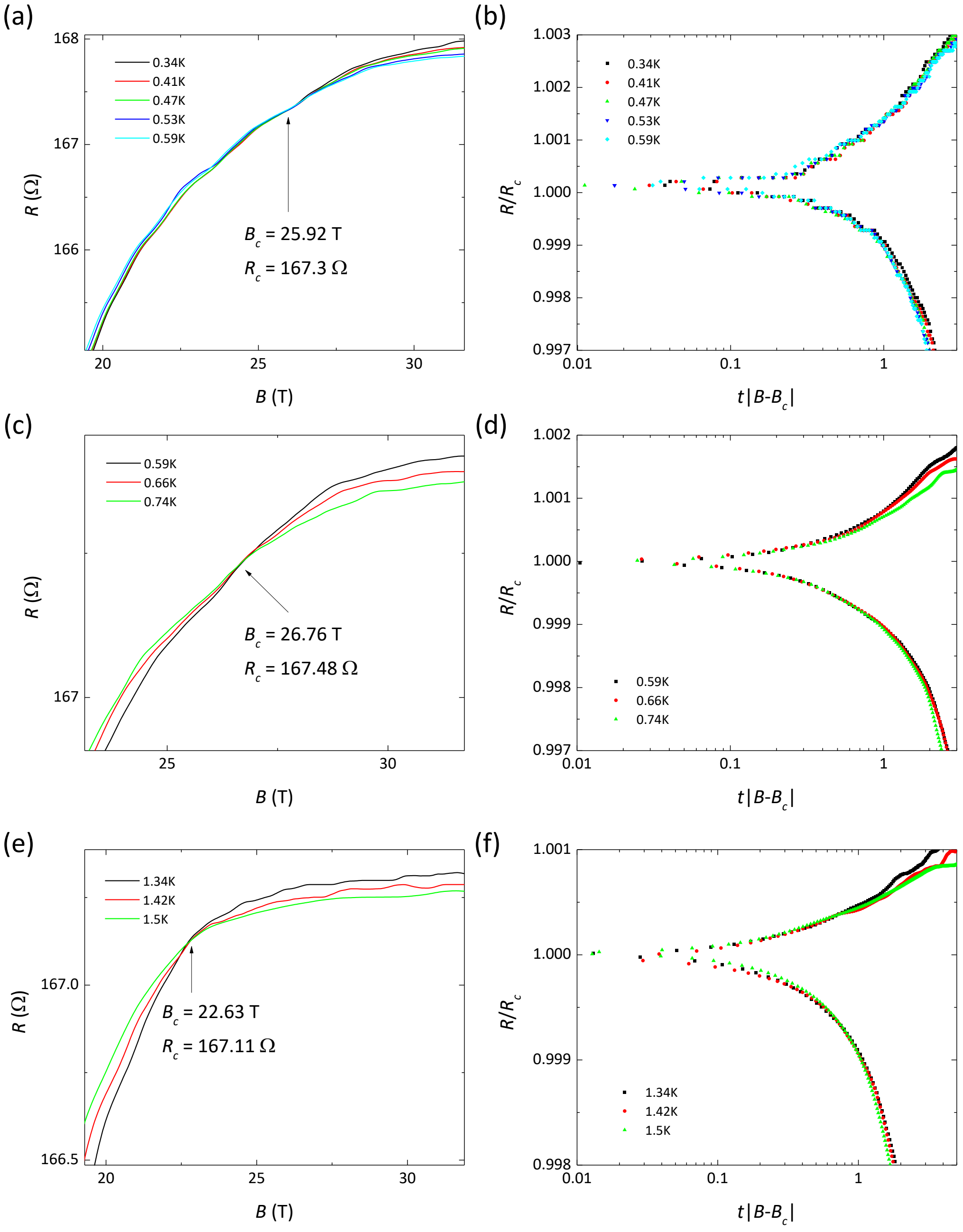

Fig. S16. FSS analysis for the 8 nm-thick sample (#09). (a) Resistance as a function of $B$ for different temperatures from $0.34$ to $0.59\,\mathrm{K}$. The crossing point is at $B_c = 25.92\,\mathrm{T}$, $R_c = 167.3\,\Omega$. (b) FSS plot of $R/R_{c1}$ as a function of scaling variable $t|B - B_c|$, with $t = T/T_0{}^{-1/z\nu}$. (c-d) Resistance as a function of $B$ and FSS plot of $R/R_{c2}$ for different temperatures from $0.59$ to $0.74$ K, respectively. (e-f) Resistance as a function of $B$ and FSS plot of $R/R_{c2}$.

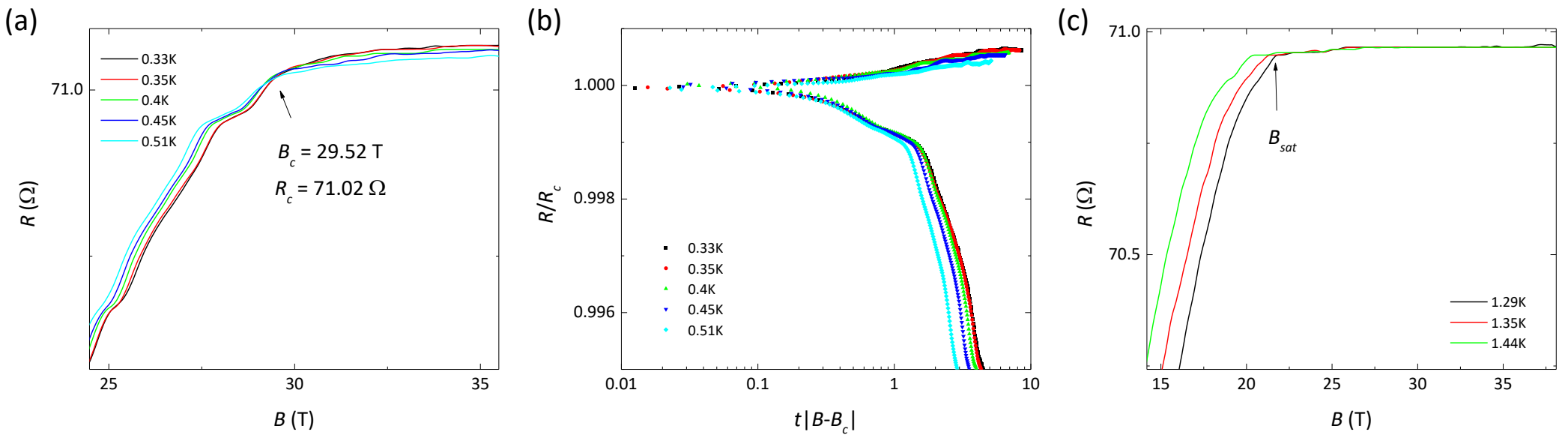

Fig. S17. Inappropriate FSS analysis for the 22 nm-thick sample (#16). (a) Resistance as a function of $B$ for different temperatures from 0.33 to 0.51 K. The crossing point is at $B_c = 29.52$ T, $R_c = 71.02\ \Omega$. (b) FSS analysis of $R/R_{c1}$ as a function of scaling variable $t|B - B_c|$, with $t = T/T_0{}^{-1/z\nu}$. Only low $t|B - B_c| < 1$ T region can be fitted, indicating the FSS analysis is not applicable. (c) Resistance as a function of $B$ for different temperatures from 1.29 to 1.44 K. $B_{sat}$ represents the magnetic field at which the resistance starts to saturate.

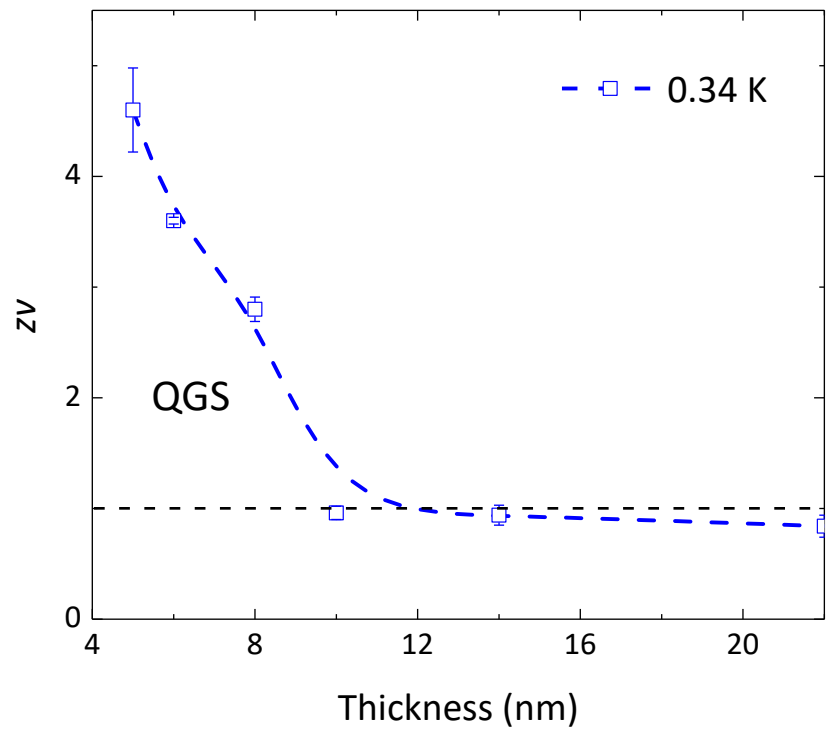

Fig. S18 | Thickness-dependent critical exponent $z\nu$ at 0.34 K. The $z\nu > 1$ region at 0.34 K indicates the QGS phase in samples thinner than 8 nm.

**Table S2 | Estimated transport parameters in typical $\beta$ − W samples (2D)**

| Sample | Thickness (nm) | $T_c$ (K) | $H_{c2\perp}$ (T) | $H_{c2\parallel}$ (T) | $\xi_{GL}$ (nm) | $n_{2D}$ (cm$^{-2}$) | $k_F$ (Å$^{-1}$) | $\tau_{so}$(fs) | $l_m$ (nm) |
|---|---|---|---|---|---|---|---|---|---|
| #01 | ~2 | 1.64 | 2.4 | 30.6 | 12 | $2.3 \times 10^{16}$ | 3.8 | 18 | 0.5 |
| #05 | ~4 | 3.30 | 4.0 | 33 | 9.1 | $8.0 \times 10^{16}$ | 7.1 | 28 | 1.6 |
| #07 | ~6 | 2.90 | 4.1 | ~28 | 9.0 | $7.6 \times 10^{16}$ | 6.9 | 34 | 3.1 |
| #10 | ~10 | 2.34 | 3.7 | 13 | 9.4 | $1.4 \times 10^{17}$ | 9.4 | $1.5 \times 10^{2}$ | 3.8 |
| #11 | ~12 | 2.30 | 3.6 | 11 | 9.6 | $1.4 \times 10^{17}$ | 9.4 | $1.8 \times 10^{2}$ | 4.4 |
| #14 | ~16 | 2.55 | 2.6 | 8.4 | 11 | $1.3 \times 10^{17}$ | 8.9 | $3.3 \times 10^{2}$ | 6.8 |
| #16 | ~22 | 1.85 | 1.5 | 4.9 | 15 | $2.5 \times 10^{17}$ | 12.5 | $7.8 \times 10^{2}$ | 8.3 |